\documentclass[%
 aip,
 amsmath,amssymb,
 reprint,%
]{revtex4-1}

\usepackage{graphicx}
\usepackage{dcolumn}
\usepackage{bm}

\usepackage[utf8]{inputenc}
\usepackage[T1]{fontenc}
\usepackage{mathptmx}
\usepackage{etoolbox}

\usepackage{textgreek}
\usepackage{chemformula}

\makeatletter
\def\@email#1#2{%
 \endgroup
 \patchcmd{\titleblock@produce}
  {\frontmatter@RRAPformat}
  {\frontmatter@RRAPformat{\produce@RRAP{*#1\href{mailto:#2}{#2}}}\frontmatter@RRAPformat}
  {}{}
}%
\makeatother

\def\iu{\ensuremath{\mathrm{i}}}
\def\du{\ensuremath{\mathrm{d}}}

\usepackage{siunitx}
  \DeclareSIUnit\angstrom{\text{\AA}}
\usepackage{textgreek}
\usepackage{braket}
\usepackage{booktabs}

\usepackage{array}
\newcommand{\PreserveBackslash}[1]{\let\temp=\\#1\let\\=\temp}
\newcolumntype{C}[1]{>{\PreserveBackslash\centering}p{#1}}
\newcolumntype{R}[1]{>{\PreserveBackslash\raggedleft}p{#1}}
\newcolumntype{L}[1]{>{\PreserveBackslash\raggedright}p{#1}}

\newcommand{\cb}{\textbf{CB}}
\newcommand{\ccc}{C\textsubscript{3}}
\newcommand{\cc}{C\textsubscript{2}}
\newcommand{\cizoa}{\mbox{\textbf{MECI-01a}}}
\newcommand{\cizob}{\mbox{\textbf{MECI-01b}}}
\newcommand{\cizoc}{\mbox{\textbf{MECI-01c}}}
\newcommand{\ciota}{\mbox{\textbf{MECI-12a}}}
\newcommand{\ciotb}{\mbox{\textbf{MECI-12b}}}

\begin{document}

\preprint{AIP/123-QED}

\title[]{A CASSCF/MRCI Trajectory Surface Hopping Simulation of the Photochemical Dynamics and the Gas Phase Ultrafast Electron Diffraction Patterns of Cyclobutanone}
\author{Xincheng Miao}
  \affiliation{Institut für Physikalische und Theoretische Chemie, Julius-Maximilians-Universität Würzburg, Emil-Fischer-Straße 42, 97074 Würzburg, Germany.}
\author{Kira Diemer}%
  \affiliation{Institut für Physikalische und Theoretische Chemie, Julius-Maximilians-Universität Würzburg, Emil-Fischer-Straße 42, 97074 Würzburg, Germany.}

\author{Roland Mitri\'{c}}
  \email{roland.mitric@uni-wuerzburg.de.}
  \affiliation{Institut für Physikalische und Theoretische Chemie, Julius-Maximilians-Universität Würzburg, Emil-Fischer-Straße 42, 97074 Würzburg, Germany.}

\date{\today}

\begin{abstract}
We present the simulation of the photochemical dynamics of cyclobutanone induced by the excitation of the 3s Rydberg state. For this purpose, we apply the complete active space self-consistent field method together with spin-orbit multireference configuration interaction singles treatment, combined with the trajectory surface hopping for inclusion of the nonadiabatic effects. The simulations were performed in the spin-adiabatic representation including nine electronic states derived from three singlet and two triplet spin-diabatic states. Our simulations reproduce the two previously observed primary dissociation channels: the \cc{} pathway yielding \ch{C2H4 + CH2CO} and the \ccc{} pathway producing \mbox{\ch{"c-" C3H6 + CO}}. In addition, two secondary products, \ch{CH2 + CO} from the \cc{} pathway and \ch{C3H6} from the \ccc{} pathway, both of them previously reported, are also observed in our simulation. We determine the ratio of the \ccc{}:\cc{} products to be about 2.8. Our findings show that most of the trajectories reach their electronic ground state within \SI{200}{\fs}, with dissociation events finished after \SI{300}{fs}. We also identify the minimum energy conical intersections that are responsible for the relaxation and provide an analysis of the photochemical reaction mechanism based on multidimensional scaling. Furthermore, we demonstrate a minimal impact of triplet states on the photodissociation mechanism within the observed timescale. In order to provide a direct link to experiments, we simulate the gas phase ultrafast electron diffraction patterns and connect their features to the underlying structural dynamics.
\end{abstract}

\maketitle

\section{\label{sec:introduction}Introduction}
Cyclobutanone (\cb{}) plays a pivotal role in the study of photochemical dynamics, since it possesses unique and interesting photochemical dynamics caused by its ring strain.\cite{lee_laser_1977} Spectroscopic studies of \cb{} with photon energies up to \SI{180}{\nm} have found two electronic transitions\cite{whitlock_electronic_1971}, with the $\mathrm{S_0} \to \mathrm{S_1}$ band having its absorption maximum at around \SI{280}{\nm}, which is assigned to the symmetry forbidden $\mathrm{n} \to \mathrm{\pi^*}$ transition \cite{hemminger_laser-excited_1973}, and the $\mathrm{S_0} \to \mathrm{S_2}$ band centered at about \SI{195}{\nm}, which is assigned to the $\mathrm{n} \to \mathrm{3s}$ Rydberg transition\cite{otoole_vacuum-ultraviolet_1991}. Early experimental photodissociation studies have established that \cb{} in the gas phase decomposes by two primary processes to yield \ch{C2H4 + CH2CO} (\cc{} pathway) and \ch{"c-" C3H6 + CO} (\ccc{} pathway) upon excitation to the S\textsubscript{1} state.\cite{benson_photochemical_1942,blacet_photochemical_1957} The \ccc{}:\cc{} product ratio was found to have complicated nonmonotonic dependence on the excitation wavelength, being 0.8 at \SI{248}{\nm}, 0.4 at \SI{313}{\nm}, 2 at \SI{326}{\nm} and as high as 7 at \SI{344}{\nm} in the gas phase, and can be wildly different in the solution phase.\cite{lee_tracer_1969,hemminger_unusual_1971,hemminger_fluorescence_1972,tang_laser_1976} A few studies covered excitation to the S\textsubscript{2} Rydberg state of \cb{}, which found no significant difference in the products compared to those obtained from the S\textsubscript{1} state and thus inferred that S\textsubscript{2} undergoes rapid internal conversion to S\textsubscript{1} and therefore does not have considerable impact on the reaction dynamics.\cite{baba_multiphoton_1984,campbell_mechanistic_1967,trentelman_193-nm_1990}

More recent experimental studies have delved into the ultrafast photodissociation mechanism of \cb{}. Notably, Diau \emph{et al.} employed fs-resolved mass spectroscopy to study the predissociation dynamics in the S\textsubscript{1} state and measure the rate constant.\cite{diau_femtosecond_1998,diau_femtochemistry_2001} Kao \emph{et al.} applied transient absorption spectroscopy to study the time-dependent S\textsubscript{1} population of \cb{}, as well as its larger-ring counterparts, cyclopentanone and cyclohexanone, in solution phase.\cite{kao_effects_2020} A comparative study of the $\mathrm{S_2}\to \mathrm{S_1}$ internal conversion amongst these cycloketones was carried out by Kuhlman \emph{et al.} using time-resolved spectroscopy.\cite{kuhlman_coherent_2012,kuhlman_pulling_2013} They have found out that only specific vibrational modes contribute significantly to this internal conversion process in \cb{}, making it faster than the analogous transitions of \cb{}'s larger homologues. 

On the theoretical front, advancements have been made in exploring the potential energy surface (PES) of \cb{}. Diau \emph{et al.} have located several key geometries on S\textsubscript{0} and S\textsubscript{1} using the complete active space self-consistent field (CASSCF) method, and could find a possible reaction pathway on S\textsubscript{1} leading to the \ccc{} products.\cite{diau_femtochemistry_2001} They have argued that with high excitation energies, the direct S\textsubscript{1} dissociation channel is much faster than the $\mathrm{S_1 \to \mathrm{T_1}}$ intersystem crossing, and therefore represents the major reaction pathway. This stands in contradiction with earlier studies, where the sole precursor for \ccc{} products is the T\textsubscript{1} state.\cite{denschlag_benzene_1968} A comprehensive PES scan of different cycloketones in S\textsubscript{1} and T\textsubscript{1}, including \cb{}, using the 
multi-state complete active space second-order perturbation theory (MS-CASPT2) method performed by Xia \emph{et al.} also leads to the conclusion that the S\textsubscript{1} state is primarily responsible for the photodissociation of \cb{}.\cite{xia_excited-state_2015} Trajectory-based nonadiabatic dynamics simulations of cyclopropanone\cite{janos_what_2023}, cyclobutanone\cite{liu_new_2016} and cyclohexanone\cite{shemesh_photochemical_2016} confirmed the ultrafast nature of the $\mathrm{S_1}\to \mathrm{S_0}$ transition in the highly strained species (cyclopropanone and cyclobutanone) with time constants of less than \SI{500}{\fs}, while in cyclohexanone, this process takes tens of picoseconds to finish, opening possibilities for intersystem crossing to affect the reaction mechanism. This rate difference was also found for the $\mathrm{S_2}\to \mathrm{S_1}$ transition using a vibronic coupling model.\cite{kuhlman_symmetry_2012}

However, to the best of our knowledge, no dynamics simulations covering the whole $\mathrm{S_2} \to \mathrm{S_0}$ relaxation process of \cb{} have been performed before and the effect of triplet states was never explicitly included in all previous dynamics simulations of cycloketones. In this work, we present a detailed investigation of the ultrafast photodissociation of cyclobutanone using CASSCF combined with multireference configuration interaction (MRCI) and including spin-orbit coupling. We perform trajectory surface hopping simulations of the photochemical dynamics within the manifold of spin-orbit coupled singlet and triplet states. In addition, gas phase ultrafast electron diffraction patterns were simulated to offer comparability with future experiments.

\section{\label{sec:methodology}Methodology}
\subsection{\label{subsec:methodology_electronic_structure}Electronic Structure}
The electronic structure of \cb{} has been described in the framework of the state-averaged complete active space self-consistent field (SA-CASSCF) method\cite{werner_quadratically_1981} including the first three singlet and two triplet states, followed by a multireference configuration interaction singles (MRCIS) treatment\cite{buenker_individualized_1974} to include dynamical correlation. The MRCIS calculations were carried out under the frozen core approximation. For a proper description of Rydberg states, the aug-cc-pVDZ basis set\cite{kendall_electron_1992} was chosen for the heavy atoms, whereas the smaller cc-pVDZ basis set\cite{dunning_gaussian_1989} was applied for the hydrogen atoms for reasons of computational efficiency. All electronic structure calculations were performed using the COLUMBUS program package.\cite{lischka_high-level_2001,lischka_columbusprogram_2011,lischka_columbus_2022}

The geometry of the ground electronic state has been optimized at the single-state CASSCF/MRCI level of theory in combination with GDIIS\cite{csaszar_geometry_1984} and natural internal coordinates\cite{fogarasi_calculation_1992}. As an active space, we selected eight electrons in eight orbitals for the CASSCF calculation and six electrons in six orbitals for the generation of reference configurations for the MRCIS calculation. Subsequently, vertical excitation energies of \cb{} at the optimized geometry were calculated using active spaces of varying size in order to determine the optimal active space for the nonadiabatic dynamics simulations. As shown in Table~\ref{tab:variation_active_space}, the excitation energies of SA-CASSCF calculations vary up to \SI{0.9}{\eV} across all active spaces, whereas those from the SA-CASSCF/MRCIS method show less variance, with the $\mathrm{S_0} \to \mathrm{S_1}$ and $\mathrm{S_0} \to \mathrm{S_2}$ transition energies of CASSCF(8,8)/MRCIS(6,6) showing the best agreement with the experimental absorption maxima at \SI{280}{\nm}\cite{hemminger_laser-excited_1973} and \SI{195}{\nm}\cite{otoole_vacuum-ultraviolet_1991}, respectively, which are also nicely reproduced by the CASSCF(6,6)/MRCIS(6,6) method. Since the smaller active space is less computationally demanding, has less convergence issues and can still achieve accurate results, we settle on this method for all subsequent electronic structure calculations in this study, henceforth labeled as SA(3/2)-CASSCF(6,6)/MRCIS. The chosen active space is comprised of the nonbonding $\mathrm{n}$ orbital of the oxygen atom, the $\mathrm{\pi}$ and $\mathrm{\pi^*}$ orbitals of the carbonyl group, the $\mathrm{\sigma}$ and $\mathrm{\sigma^*}$ orbitals on the bonds between the carbonyl carbon and the \textalpha-carbons, as well as the $\mathrm{3s}$ Rydberg orbital. A visualization of these orbitals can be found in Fig. S1 and S2.

\begin{table}[htbp]
    \centering
    \caption{Vertical excitation energies in \si{eV} of \cb{} at the equilibrium geometrycalculated for different active spaces at the SA(3/2)-CASSCF and the SA(3/2)-CASSCF/MRCIS level of theory, as well as the experimental absorption maxima from Ref. \onlinecite{hemminger_laser-excited_1973} and \onlinecite{otoole_vacuum-ultraviolet_1991}. The active spaces are denoted by $(n_{\mathrm{electron}}, n_{\mathrm{orbital}})$. The column in boldface indicates the method used for dynamics simulations.}
    \label{tab:variation_active_space}
    \begin{tabular}{ccccccc}
        \toprule[1.0pt]
        MCSCF & (6,6) & \multicolumn{2}{c}{(8,8)} & (10,8) & (8,7) & exp. \\ \midrule[1.0pt]
        S\textsubscript{1} & 4.26 & \multicolumn{2}{c}{4.73} & 3.94 & 4.38 & 4.4 \\
        S\textsubscript{2} & 7.06 & \multicolumn{2}{c}{7.20} & 6.80 & 7.17 & 6.2 \\
        T\textsubscript{1} & 3.96 & \multicolumn{2}{c}{4.37} & 3.70 & 4.05 & - \\
        T\textsubscript{2} & 5.67 & \multicolumn{2}{c}{6.12} & 5.28 & 5.76 & - \\ 
        \midrule[1.0pt]
        MRCI & \textbf{(6,6)} & (8,8) & (6,6) & (10,8) & (8,7) & exp. \\ \midrule[1.0pt]
        S\textsubscript{1} & \textbf{4.37} & 4.24 & 4.41 & 4.16 & 4.40 & 4.4 \\
        S\textsubscript{2} & \textbf{6.27} & 6.24 & 6.30 & 5.91 & 6.28 & 6.2 \\
        T\textsubscript{1} & \textbf{4.09} & 3.98 & 4.13 & 3.81 & 4.09 & - \\
        T\textsubscript{2} & \textbf{6.14} & 5.97 & 6.15 & 5.82 & 6.12 & - \\
        \bottomrule[1.0pt]
    \end{tabular}
\end{table}


Since we explicitly include singlet-triplet transitions in our dynamics simulations, spin-orbit coupling was taken into account in the MRCIS calculation using the effective one-electron approach.\cite{yabushita_spinorbit_1999,mai_perturbational_2014}
A comparison of the CI wavefunctions and energies obtained by a variational and a quasi-degenerate perturbative spin-orbit CI calculation at the optimized \cb{} geometry shows minimal difference between them: The coefficients for spin-diabatic states differ by no more than 0.002 for the singlets and 0.2 for the triplets, while the energy differences are less than \SI{e-4}{\eV} across all included states. Thus, in the dynamics simulations, we employ the quasi-degenerate perturbation theory (QDPT) to include spin-orbit coupling effects.

\subsection{\label{subsec:methodology_nonadiabatic_dynamics}Nonadiabatic Dynamics}

The nonadiabatic dynamics simulations were performed using Tully's fewest-switches surface hopping\cite{tully_molecular_1990} procedure combined with the CASSCF/MRCI electronic structure method. Such combination of methods has been demonstrated to be able to accurately reproduce the experimental pump-probe spectra and reveal the mechanism of the photochemical reactions involving conical intersections.\cite{mitric_ab_2006,karashima_ultrafast_2021,karashima_ultrafast_2023} In this approach, the classical Newtonian equations are used to describe the nuclear motion, while the electronic dynamics is addressed through the time-dependent Schrödinger equation, given by
\begin{equation}
    \label{eq:sh_equation}
    \iu \hbar \dot{c}_i(t) = \sum_j V_{ij}(\vec{R}(t))c_j(t) - \iu \hbar \sum_j D_{ij} (\vec{R}(t))c_j(t)\,,
\end{equation}
where $c_i(t)$ represents the expansion coefficient of the electronic wavefunction in terms of the $i$-th electronic state and $V_{ij}$ is the matrix element of the electronic Hamiltonian with respect to the electronic states $i$ and $j$, whereas $D_{ij}$ stands for the scalar nonadiabatic coupling between the electronic states $i$ and $j$, defined as
\begin{equation}
D_{ij} (\vec{R}(t)) = \underbrace{\braket{\psi_i (\vec{R}(t))|\vec{\nabla}_R|\psi_j (\vec{R}(t))}}_{\vec{d}_{ij}} \cdot \frac{\du  \vec{R}(t)}{\du t}\,.
\end{equation}

In our work, the generalized hopping probability defined in ref. \onlinecite{lisinetskaya_simulation_2011} was applied to describe the population transfer, formulated using the electronic state populations $\rho_{ii} = c_i^*c_i$. The transition probability from state $i$ to state $j$ is expressed as the product of two factors: the probability of depopulating state $i$ ($P_{i,\text{depopulation}}$) and the probability of populating state $j$ ($P_{j,\text{population}}$), calculated as follows:
\begin{subequations}
\label{eq:sh_probabilities}
\begin{align}
P_{i,\text{depopulation}} &= \Theta(-\dot{\rho}_{ii}) \frac{-\dot{\rho}_{ii}}{\rho_{ii}} \Delta t \\
P_{j,\text{population}} &= \frac{\Theta(\dot{\rho}_{jj})\dot{\rho}_{jj}}{\sum_k \Theta(\dot{\rho}_{kk})\dot{\rho}_{kk}}\,.\,
\end{align}
\end{subequations}
where $\Theta$ is the Heaviside step function.

In the electronic structure calculations, the total Hamiltonian $H^{\mathrm{tot}}$ is expressed as the sum of the molecular Coulomb Hamiltonian (MCH) and the spin-orbit (SO) Hamiltonian\cite{pederzoli_new_2017}, i.e.
\begin{equation}
    \label{eq:total_hamiltonian}
    H^{\mathrm{tot}} = H^{\mathrm{MCH}} + H^{\mathrm{SO}}\,.
\end{equation}
In the two most common approaches to incorporate spin-orbit coupling into surface hopping dynamics, the electronic wavefunction is expanded either in eigenstates of $H^{\mathrm{MCH}}$ (spin-diabatic basis $\ket{\phi_i}$) or eigenstates of $H^{\mathrm{tot}}$ (spin-adiabatic basis $\ket{\bar{\phi}_i}$)\cite{granucci_surface_2012}, with corresponding expansion coefficients $c_i(t)$ and $\bar{c}_i(t)$. The former approach, however, was shown by Granucci \emph{et al.} to be incorrect in general and to lead to unphysical behavior in all but the simplest cases.\cite{granucci_surface_2012} Therefore, the spin-adiabatic basis should be preferably applied. 

Although the spin-adiabatic approach is compatible with Eq. \eqref{eq:sh_equation}, gradients and nonadiabatic couplings of spin-adiabatic states are required, which are not available in common quantum chemistry software packages. In addition, when the spin-orbit coupling is small, the crossings between different multiplets become nearly unavoided, which requires either an extremely small step size for the integration of Eq. \eqref{eq:sh_equation} or extra treatments, and will otherwise lead to unphysical transitions.\cite{granucci_direct_2001,fabiano_implementation_2008} The first issue can be circumvented in the framework of quasi-degenerate perturbation theory (QDPT). Because the spin-adiabatic states can be expressed as linear combinations of spin-diabatic states, a unitary transformation $\bm{U}$ can be defined such that $ \ket{\bar{\phi}_i} = \sum_j U_{ji}^* \ket{\phi_j}$, leading to the equivalent equation of motion for the coefficients $\vec{\bar{c}} = \bm{U}^\dag \vec{c}$. By inserting this transformation into Eq. \eqref{eq:sh_equation}, the spin-adiabatic electronic Schrödinger equation can be obtained as
\begin{equation}
    \label{eq:sad_sh_equation}
    \iu \hbar \dot{\bar{c}}_i = E_i^{\mathrm{tot}} \bar{c}_i - \iu \hbar \sum_j \bar{D}_{ij} \bar{c}_j - \iu \hbar \sum_{jk} U_{ji}^* \dot{U}_{jk} \bar{c}_k\,,
\end{equation}
where $E_i^{\mathrm{tot}}$ is the energy of the $i$-th spin-adiabatic state and $\bar{\bm{D}} = \bm{U}^{\dag} \bm{D} \bm{U}$ are the transformed scalar couplings. The second issue that arises is connected with the nonuniqueness of $\bm{U}$. The phase of each eigenstate is arbitrary and the degenerate states can be arbitrarily mixed. In order to address this issue, several approaches have been developed previously.\cite{mai_sharc_2014,mai_general_2015,pederzoli_new_2017} 

In this work, we use the singular value decomposition (SVD) propagator method developed by Pederzoli and Pittner\cite{pederzoli_new_2017}, which will be briefly outlined here. Readers interested in more details are referred to their original publication (Ref. \onlinecite{pederzoli_new_2017}). We start by defining general basis functions $\ket{\tilde{\phi}_i} = \sum_j \tilde{U}_{ji}^* \ket{\phi_j}$ ($\vec{\tilde{c}} = \tilde{\bm{U}}^\dag \vec{c}$), which do not necessarily diagonalize $H^{\mathrm{tot}}$. This leads to the generalized surface hopping equation
\begin{equation}
    \label{eq:generalized_sh_equation}
    \iu \hbar \dot{\tilde{c}}_i = \sum_j \tilde{H}_{ij} \tilde{c}_j - \iu \hbar \sum_j \tilde{D}_{ij} \tilde{c}_j - \iu \hbar \sum_{jk} \tilde{U}_{ji}^* \dot{\tilde{U}}_{jk} \tilde{c}_k
\end{equation}
with the transformed scalar coupling matrix $\tilde{\bm{D}} = \tilde{\bm{U}}^{\dag} \bm{D} \tilde{\bm{U}}$ and the transformed Hamiltonian $\tilde{\bm{H}} = \tilde{\bm{U}}^{\dag} \bm{H}^{\mathrm{tot}} \tilde{\bm{U}}$. The time derivative of the transformation matrix $\dot{\tilde{\bm{U}}}$ is computed using the finite difference approximation as
\begin{equation}
    \label{eq:u_tilde_dot}
    \dot{\tilde{\bm{U}}}(t) = \frac{\bm{U}(t + \Delta t) - \bm{U}'(t)}{\Delta t}\,,
\end{equation}
where the matrix $\bm{U}'(t)$ is obtained from a linear transformation of $\bm{U}$ that aligns all phases. As shown by Pederzoli and Pittner, this can be realized by
\begin{equation}
    \label{eq:u_prime}
    \bm{U}'(t) = \bm{U}(t) \bm{u} \bm{v}^\dag
\end{equation}
with $\bm{u}$ and $\bm{v}$ being matrices of the left and right singular vectors of the overlap matrix $\bm{S}$, which is defined as
\begin{equation}
    \label{eq:transform_matrix_overlap}
    S_{ij} = \begin{cases}
        0 & \sum_k U_{ki}^*(t) U_{kj}(t + \Delta t) < \epsilon \\
        \sum_k U_{ki}^*(t) U_{kj}(t + \Delta t) & \text{otherwise}
    \end{cases}\,.
\end{equation}
The threshold $\epsilon$ to zero matrix elements corresponding to nondegenerate states is chosen to be \mbox{$0.0001\ E_\mathrm{h}$}.

The trajectories are propagated with the electronic wavefunction being expanded using the coefficients $\bar{c}_i$. At each time step, the transformation $\bar{c}_i(t) \to \tilde{c}_i(t)$ is performed to solve the Eq. \eqref{eq:generalized_sh_equation} and obtain $\tilde{c}_i(t + \Delta t)$. Inverse transformation $\tilde{c}_i(t + \Delta t) \to \bar{c}_i(t + \Delta t)$ delivers the new spin-adiabatic coefficients.
The hopping probabilities are calculated using Eq. \eqref{eq:sh_probabilities} but with spin-adiabatic populations $\bar{\rho}_{ii} = \bar{c}_i^* \bar{c}_i$. Because the nuclear degrees of freedom are propagated on the spin-adiabatic surfaces, the corresponding nuclear gradients $\vec{\nabla}_R E_i^{\mathrm{tot}}$ are approximated by the following expression:
\begin{equation}
    \label{eq:sad_grad_approx}
    \vec{\nabla}_R E_i^{\mathrm{tot}} = \sum_{jk} U_{ji}^* \left[ \delta_{jk} \vec{\nabla}_R E_j^{\mathrm{MCH}} + (E_k^{\mathrm{MCH}} - E_j^{\mathrm{MCH}}) \vec{d}_{jk} \right] U_{ki}\,,
\end{equation}
assuming the matrix elements of $\nabla_R H^{\mathrm{SO}}$ are small.\cite{mai_general_2015}

For our dynamics simulations, the electronic Schrödinger equation was solved at the SA(3/2)-CASSCF(6,6)/MRCIS level of theory with 3 singlet and 2 triplet roots coupled \emph{via} QDPT. The initial structures were generated by sampling a harmonic canonical Wigner distribution\cite{wigner_quantum_1932,bonacic-koutecky_theoretical_2005} at \SI{50}{\K}. In total, 52 trajectories were launched from the bright spin-adiabatic state with dominant S\textsubscript{2} contribution ($>\SI{99.9}{\percent}$), and propagated for \SI{360}{\fs} with a time step of \SI{0.2}{\fs}. Some trajectories faced convergence problems for the S\textsubscript{2} and T\textsubscript{2} states after dissociation, and were continued at the SA(2/1)-CASSCF(6,6)/MRCIS level of theory. 

\subsection{\label{subsec:methodology_electron_diffraction}Electron Diffraction Patterns}
In order to establish a direct connection to experiment, the gas phase electron diffraction (GED) patterns were calculated using the independent atom model (IAM) within the framework of first Born approximation.\cite{centurion_ultrafast_2022} In this model, atoms do not interact with each other and can therefore be described using precalculated atomic form factors. The elastic scattering intensity can thus be calculated as
\begin{equation}
    \label{eq:ged_intensity_angular}
    I(\vec{s}) = \sum_{i} f_i^2(s) + \sum_{i} \sum_{j \neq i} f_i(s) f_j(s) \exp (\iu \vec{s} \cdot (\vec{r}_i - \vec{r}_j))\,,
\end{equation}
where $f_i(s)$ is the tabulated atomic form factor of the $i$-th atom and $\vec{r}_i$ is the position vector of the $i$-th atom obtained from the nonadiabatic dynamics simulations.\cite{brockway_electron_1936} By assuming the molecules are randomly oriented, a rotational averaging can be performed to obtain 
\begin{equation}
    \label{eq:ged_intensity}
    I(s) = \underbrace{\sum_{i} f_i^2(s)}_{I_{\mathrm{at}}(s)} + \underbrace{\sum_{i} \sum_{j \neq i} f_i(s) f_j(s) \frac{\sin (s r_{ij})}{s r_{ij}}}_{I_{\mathrm{mol}}(s)}
\end{equation}
with $r_{ij} = \|\vec{r}_i - \vec{r}_j\|$. 

Since we want to focus on the time evolution of the diffraction pattern, a more suitable measure is the modified time-dependent difference scattering intensity $\Delta sM(s;t)$, which can be calculated as
\begin{equation}
    \label{eq:ged_difference_signal}
    \Delta sM(s;t) = s \frac{I_{\mathrm{mol}}(s;t) - I_{\mathrm{mol}}(s;0)}{I_{\mathrm{at}}(s)}
\end{equation}
using the IAM, where $I_{\mathrm{mol}}(s;0)$ is evaluated on the optimized ground state geometry. Since the inelastic scattering signal is independent of the molecular geometry within the IAM, it does not have to be accounted for explicitly in the difference signal.\cite{centurion_ultrafast_2022} 

For a more intuitive interpretation of the diffraction pattern, the difference signal is transformed back into real space to obtain the difference pair distribution function 
\begin{equation}
    \label{eq:ged_difference_pdf}
    \Delta P(r;t) \approx r \cdot \int_{s_\mathrm{min}}^{s_\mathrm{max}} \Delta sM (s;t) \sin(sr) \exp(-\alpha s^2) \du s\,,
\end{equation}
where the damping factor $\exp(-\alpha s^2)$ applies Gaussian smoothing in real space to avoid numerical artifacts of the sine transform.

\section{\label{sec:results_and_discussion}Results and Discussion}

\subsection{\label{subsec:results_electronic_structures}Structural and Electronic Properties of \cb{}}
\begin{figure}[htbp]
    \centering
    \includegraphics[width=\linewidth]{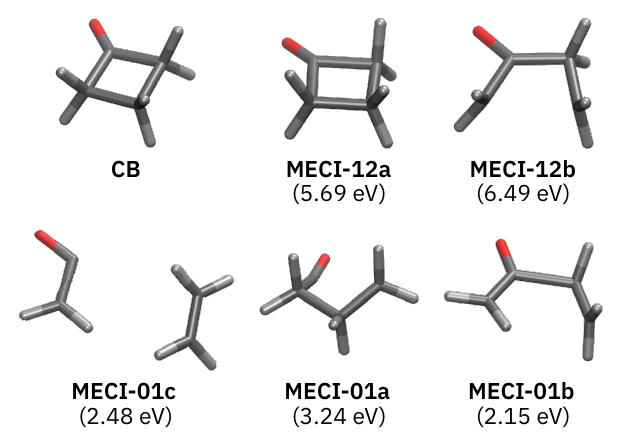}
    \caption{Optimized ground state geometry of \cb{} at the single state CASSCF(8,8)/MRCIS(6,6) level of theory and optimized minimal energy conical intersections (MECIs) at the SA(3/2)-CASSCF(6,6)/MRCIS level of theory. The energies of MECIs relative to the optimized \cb{} structure are shown in parentheses.}
    \label{fig:geometries}
\end{figure}

In this section, we characterize relevant geometries and electronic states involved in the photochemistry of \cb{}. Previously, three minimum-energy conical intersections (MECI) between S\textsubscript{1} and S\textsubscript{0} enabling the $\mathrm{S_1} \to \mathrm{S_0}$ internal conversion (\cizoa{}, \cizob{} and \cizoc{})  have already been reported by Liu and Fang.\cite{liu_new_2016} These MECIs were re-optimized by means of SA(3/2)-CASSCF(6,6)/MRCIS in combination with the Lagrange-Newton method\cite{manaa_intersection_1993,dallos_analytic_2004}, and are depicted in Fig. \ref{fig:geometries}. Liu and Fang have found that the \cc{} pathway was accessible from all three MECIs, whereas the \ccc{} pathway could only be reached through one of them, here denoted as \cizoa{}. In this structure, the bond between the carbonyl carbon (C\textsubscript{O}) and one \textalpha-carbon (C\textsubscript{\textalpha}) is broken, while the remaining two MECIs, labeled as \cizob{} and \cizoc{}, differ from the optimized ground state geometry by a broken \ch{C\textsubscript{\textalpha}-C\textsubscript{\textbeta{}}} bond and a partially or fully cleaved \ch{C\textsubscript{O}-C\textsubscript{\textalpha{}}} bond. The relative energies of \cizoa{}, \cizob{} and \cizoc{} with respect to \cb{} in the ground state are 3.24, 2.15 and \SI{2.48}{\eV}, respectively.

To explore the dynamics originating from the S\textsubscript{2} state of \cb{}, the consideration of MECIs between S\textsubscript{2} and S\textsubscript{1} becomes also important. We identified two such MECIs, termed as \ciota{} and \ciotb{}, which were optimized at the SA(3/2)-CASSCF(6,6)/MRCIS level of theory. The latter contains a cleaved \ch{C\textsubscript{\textalpha}-C\textsubscript{\textbeta{}}} bond, resembling \cizob{}, while the former is very similar to the ground state geometry of \cb{}. The relative energies of \ciota{} and \ciotb{} are 5.69 and \SI{6.49}{\eV}, respectively. These two MECIs are depicted in the upper part of Fig. \ref{fig:geometries}.

\begin{figure}[htbp]
    \centering
    \includegraphics[width=\linewidth]{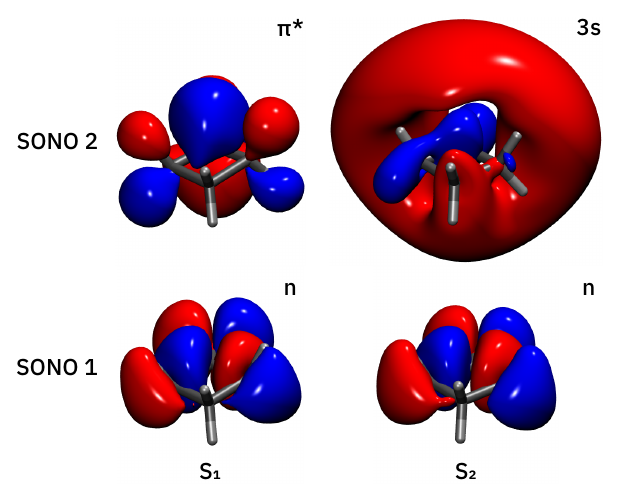}
    \caption{Nearly singly occupied natural orbitals (SONOs) from the SA(3/2)-CASSCF(8,8)/MRCI(6,6) calculation at the optimized ground state geometry of \cb{} for the excited states S\textsubscript{1} and S\textsubscript{2}. An isovalue of 0.02 was chosen. The character of each SONO is given in its upper right corner.}
    \label{fig:ci_sono}
\end{figure}

To characterize the S\textsubscript{1} and S\textsubscript{2} states of \cb{}, we have calculated the oscillator strengths and natural orbitals for them in addition to their vertical excitation energies. The results agree with the experimental findings that the $\mathrm{S_0} \to \mathrm{S_1}$ absorption corresponds to the symmetry forbidden $\mathrm{n} \to \mathrm{\pi^*}$ transition and to the $\mathrm{S_0} \to \mathrm{S_2}$ process being the $\mathrm{n} \to \mathrm{3s}$ Rydberg transition, as shown by the nearly singly occupied natural orbitals (SONOs) of both states depicted in Fig. \ref{fig:ci_sono}.

\begin{figure}[htbp]
    \centering
    \includegraphics[width=\linewidth]{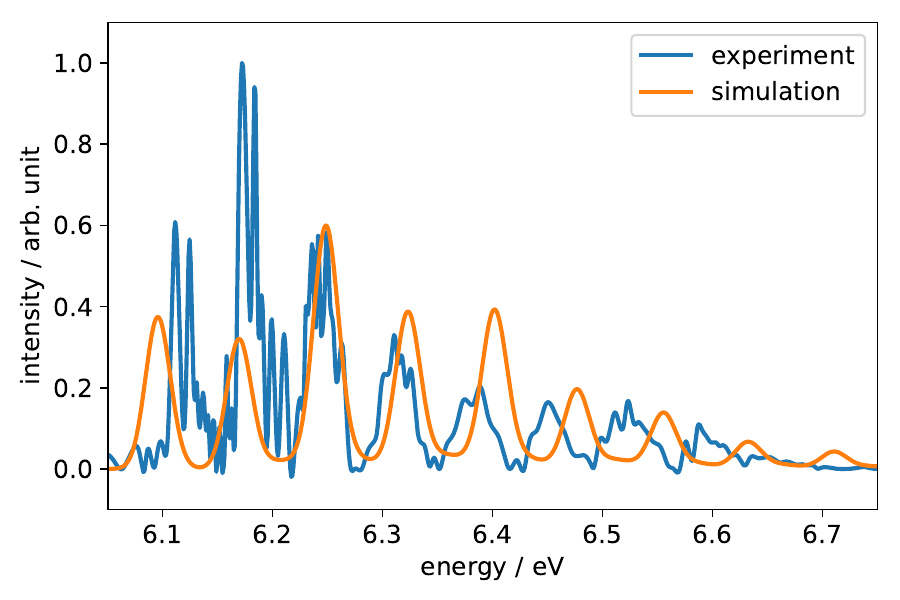}
    \caption{Simulated vibrationally resolved absorption spectrum of the S\textsubscript{2} state of \cb{} using the vertical Hessian model. The simulated stick spectrum was blueshifted by \SI{0.13}{\eV} and broadened with a Gaussian profile with the width of $\sigma=\SI{0.01}{\eV}$. The experimental absorption spectrum plotted for comparison was reproduced from Ref. \onlinecite{otoole_vacuum-ultraviolet_1991} with additional background removal using the improved asymmetric least squares algorithm\cite{he_baseline_2014} with the parameters $\lambda = 100$ and $p = 0.1$.}
    \label{fig:vh_spec}
\end{figure}

We have also calculated the vibrationally resolved absorption spectrum of S\textsubscript{2} using the vertical Hessian model\cite{ferrer_comparison_2012} as implemented in the FCclasses3 program package\cite{santoro_effective_2007,cerezo_fcclasses3_2023,santoro_fcclasses3_2023}. The simulated spectrum was blueshifted by \SI{0.13}{\eV} and broadened with a Gaussian profile of $\sigma=\SI{0.01}{\eV}$ to allow a better comparison with the experimental spectrum from Ref. \onlinecite{otoole_vacuum-ultraviolet_1991} as shown in Fig. \ref{fig:vh_spec}. Apart from the intensity of the second peak and the frequency mismatch at higher energies due to the harmonic approximation, the experimental spectrum is very well reproduced, demonstrating the accuracy of the employed electronic structure method. The vibronic structure in Fig. \ref{fig:vh_spec} is caused by the progression of the carbonyl stretching ($\nu_{12} = \SI{154}{\milli\eV}$) and its combination with the out-of-plane wagging of the carbonyl group ($\nu_{4} = \SI{75}{\milli\eV}$). Both vibrational modes are depicted in Fig. S3.

\subsection{\label{subsec:results_nonadiabatic_dynamics}Nonadiabatic Dynamics Simulations of Photochemistry}
\begin{figure}
    \centering
    \includegraphics[width=\linewidth]{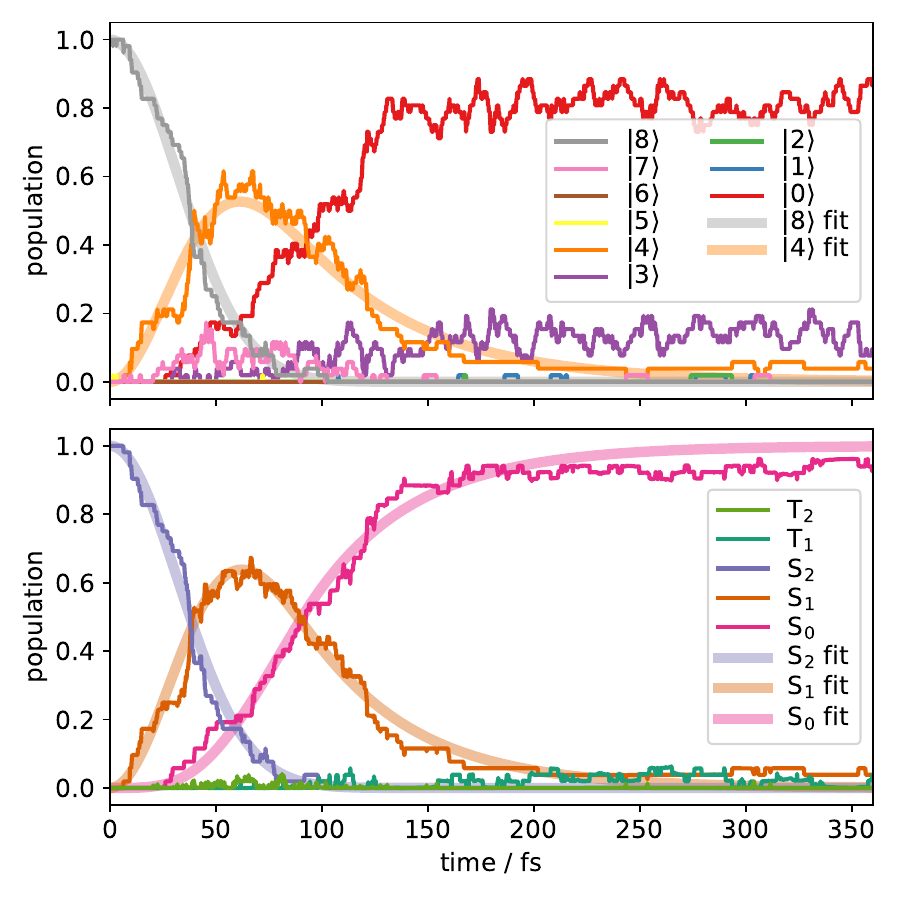}
    \caption{Time evolution of the populations in the spin-adiabatic states (top) and the spin-diabatic states (bottom). The transparent thick curves are fits according to the rate model described in Section S5 of the supporting information.}
    \label{fig:population}
\end{figure}

The time evolution of the populations in the spin-adiabatic as well as the spin-diabatic states obtained from trajectory surface hopping dynamics are shown in Fig. \ref{fig:population}. As can be seen, the bright spin-adiabatic state $\ket{8}$ is almost a pure spin-diabatic S\textsubscript{2} state, and the same holds true for the spin-adiabatic state $\ket{4}$ and the spin-diabatic state S\textsubscript{1}, indicating almost no mixing between different multiplets. This is further confirmed by the very low population in the triplet diabatic manifold. Almost all the population there is caused by the finite step size used in the surface hopping dynamics, which leads to delayed hops at trivial crossings. The deviation of $\ket{0}$ from S\textsubscript{0} shows that some biradical structures are formed through the photodissociation.

During the decay of the $\ket{8}$ (S\textsubscript{2} state), the $\ket{4}$ (S\textsubscript{1} state) is rapidly populated and reaches a maximum population of about 0.6 after \textit{ca.} \SI{70}{\fs}. Afterwards, the population is funneled into the states $\ket{3}$ and $\ket{0}$, which translate to S\textsubscript{0} with very little T\textsubscript{1} character in the spin-diabatic picture. The time evolution of the energies for all these states across different trajectories can be found in Fig. S4 and S5. 

To quantify the lifetime of the excited states, the populations of spin-adiabatic $\ket{8}$ and $\ket{4}$ states, as well as those of spin-diabatic S\textsubscript{2}, S\textsubscript{1} and S\textsubscript{0} states were fitted using the rate model described in detail in Section S5 of the supporting information. The fitted populations are shown as thick curves in Fig. \ref{fig:population}. Because the rate is a function of time in our model, we defined a mean lifetime $\bar{\tau}$ in Eq. (S.4) and (S.5), which captures the overall behavior of a decay path. For the states $\ket{8}$ and S\textsubscript{2}, their mean lifetimes were calculated to be around \SI{25}{\fs}. A lifetime of about \SI{56}{\fs} was obtained for $\ket{4}$ and S\textsubscript{1}. Using this fit, the half-life of $\ket{8}$ and S\textsubscript{2} was determined to be around \SI{38}{\fs}. 

Amongst all 52 trajectories, 13 followed the \cc{} pathway and 36 followed the \ccc{} pathway, while 1 trajectory remains as hot \cb{} in its ground state and 2 trajectories are trapped as \cb{} in the S\textsubscript{1} state. Therefore, we obtain for the 50 fully relaxed trajectories the following branching ratios: \SI{26}{\percent} (\SIrange{15}{40}{\percent}) \cc{}, \SI{72}{\percent} (\SIrange{58}{84}{\percent}) \ccc{} and \SI{2}{\percent} (\SIrange{0}{11}{\percent}) \cb{}, where the ranges in parentheses show the \SI{95}{\percent} confidence interval by assuming a binomial distribution. This leads to a \ccc{}:\cc{} ratio of 2.8. Out of the 13 trajectories following the \cc{} pathway, 1 of them dissociated on the S\textsubscript{1} surface and further reacted to \ch{CH2} and \ch{CO}. For the 36 trajectories following the \ccc{} pathway, because the \ch{C2H6} fragment is very hot, a ring closure to cyclopropane does not always happen in the short time scale in our simulation and a majority of these trajectories oscillates between distorted cyclopropane and trimethylene biradical. Out of these trajectories, 2 underwent a hydrogen shift while being trimethylene biradical, leading to the formation of propylene. Snapshots from four representative trajectories resulting in the discussed products are shown in Fig. \ref{fig:traj_snapshots}. We also provide movies of these four trajectories in the supplementary material.

\begin{figure*}[htbp]
    \centering
    \includegraphics[width=\linewidth]{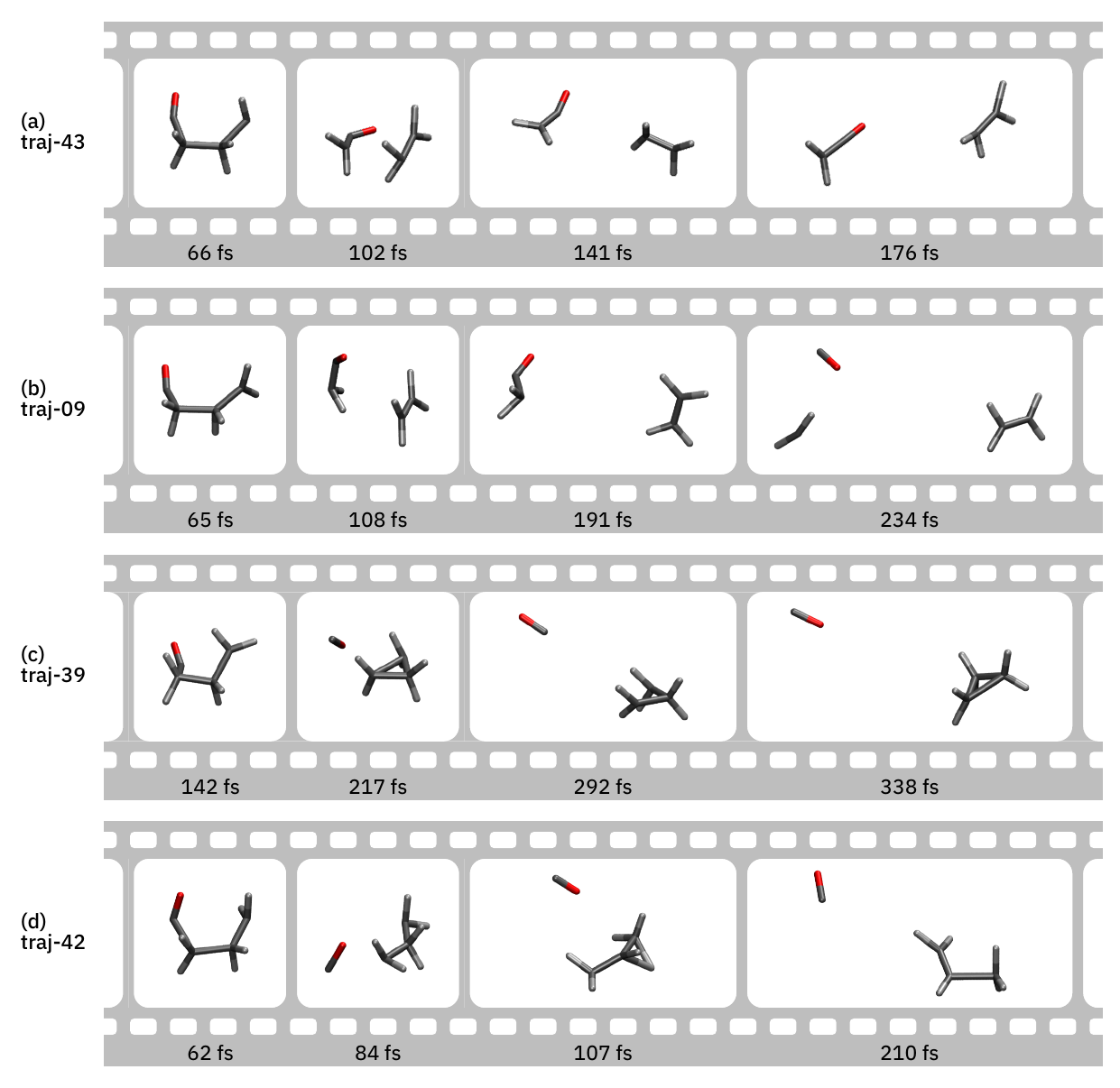}
    \caption{Snapshots from representative trajectories. (a) \cc{} pathway producing ethenone. (b) \cc{} pathway producing methylene and carbon monoxide. (c) \ccc{} pathway producing cyclobutanone. (d) \ccc{} pathway producing propylene.}
    \label{fig:traj_snapshots}
\end{figure*}

Additionally, we performed nonadiabatic dynamics simulations including only the 3 lowest singlet states using the same 52 initial conditions to see if triplet states have any significant impact on the product ratio. This simulation led to 12 trajectories following the \cc{} pathway and 35 trajectories following the \ccc{} pathway, hardly different from the results obtained by the spin-orbit coupled simulation. The exact photoproducts of each trajectory from the spin-orbit coupled dynamics as well as the pure singlet dynamics can be found in Tab. S2.

\subsection{Multidimensional Scaling Analysis of the Reaction Mechanism}

\begin{figure*}[htbp]
    \centering
    \includegraphics[width=\linewidth]{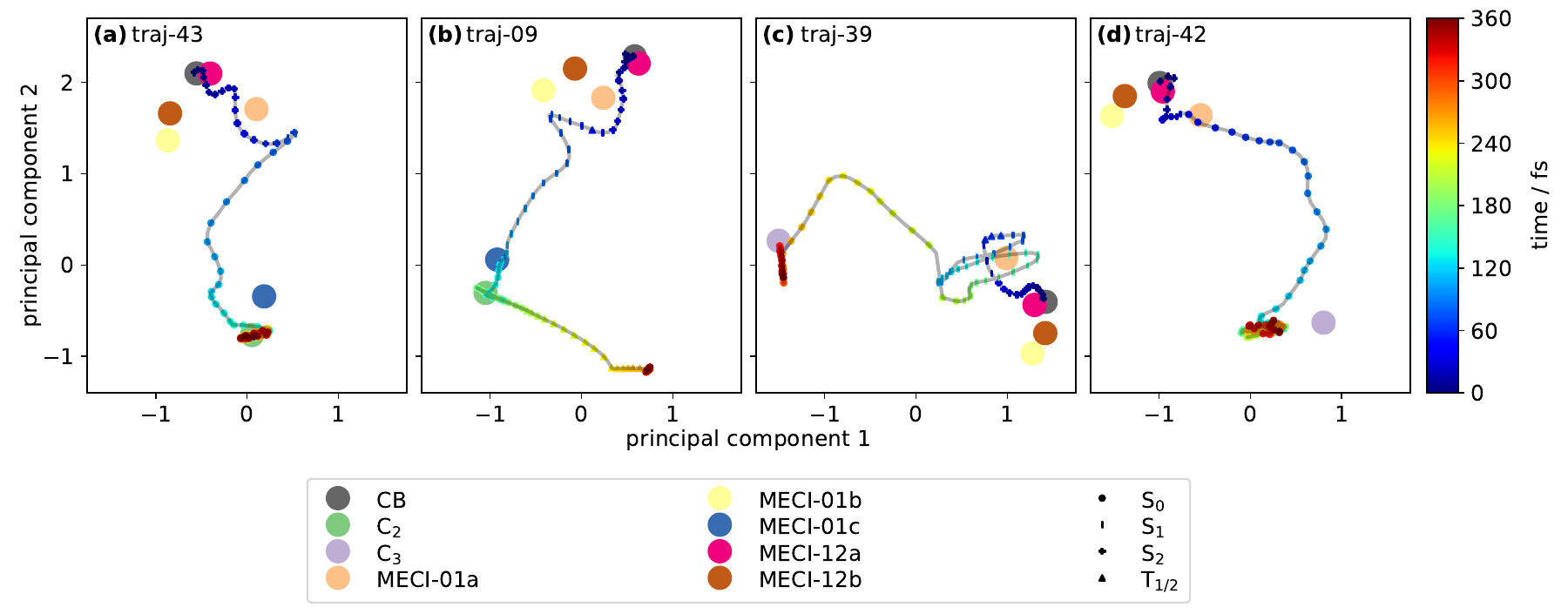}
    \caption{Path of selected trajectories embedded into two-dimensional space using multidimensional scaling. The color of markers show the time progress and their shapes denote the current spin-diabatic state of the trajectory. For clarity, only every second marker is drawn. The large circles of different colors represent reference structures. (a) \cc{} pathway producing ethenone. (b) \cc{} pathway producing methylene and carbon monoxide. (c) \ccc{} pathway producing cyclobutanone. (d) \ccc{} pathway producing propylene.}
    \label{fig:selected_mds}
\end{figure*}

To visualize and understand the reaction mechanism, we constructed a translationally and rotationally invariant representation vector $p$ for each molecule and defined a dissimilarity measure $d_{IJ}$ between pairs of molecules $I$ and $J$ as a weighted norm for the difference vector, which preserves some symmetry in the \cb{} molecule. The exact definitions of $p$ and $d$ can be found in Section S7 of the supporting information. We then performed multidimensional scaling (MDS) analysis as implemented in the scikit-learn module\cite{pedregosa_scikit-learn_2011} for 181 equidistantly selected geometries from each trajectory along with several reference structures. This allowed us to visualize and analyze the time evolution of the trajectories in configurational space. The two-dimensional projections of four representative trajectories are depicted in Fig. \ref{fig:selected_mds} as gray lines. The reference structures are represented as large circles of different colors. 

\begin{figure}[htbp]
    \centering
    \includegraphics[width=\linewidth]{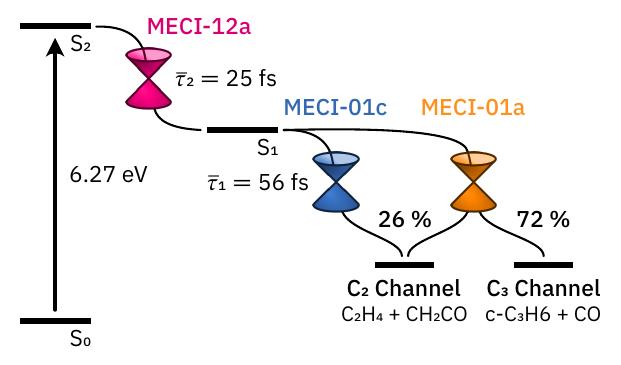}
    \caption{A schematic representation of the reaction mechanism derived from the nonadiabatic dynamics simulations.}
    \label{fig:reaction_scheme}
\end{figure}

The panels (a) and (b) in Fig. \ref{fig:selected_mds} display two trajectories following the \cc{} pathway. The dissociation happens relatively quickly at around \SI{100}{\fs} for both of them, shortly after their transition through \ciota{}. Afterwards, the trajectory in (a) relaxes into the ground state \emph{via} \cizoa{}, while the one shown in (b) passes through \cizoc{}. In Fig. \ref{fig:selected_mds}b, it is also possible to see the further dissociation after its conversion into \cc{} products. The panels (c) and (d) in Fig. \ref{fig:selected_mds} show two trajectories following the \ccc{} pathway. Both of them reached the ground state through \ciota{} and \cizoa{}, although the one (c) dissociates later at around \SI{200}{\fs} compared to the dissociation of the trajectory in (d), which happens at roughly \SI{100}{\fs}. An analysis of all trajectories in this manner shows that the $\mathrm{S_2}\to \mathrm{S_1}$ internal conversion is always realized by the \ciota{} funnel, while the $\mathrm{S_1}\to \mathrm{S_0}$ transition through \cizoc{} leads to \cc{} products, and that through \cizoa{} leads to \cc{} as well as \ccc{} products. This agrees with the findings of Liu and Fang.\cite{liu_new_2016} A schematic representation of the reaction mechanism is shown in Fig. \ref{fig:reaction_scheme}

However, the reported \cizob{} funnel in Ref. \onlinecite{liu_new_2016} could not be identified in our simulations. In addition, we could not find any trajectories relaxing through \ciotb{}. This could be explained by \ciota{} being structurally very similar to the Franck-Condon structure of \cb{}, making it easily accessible upon excitation. Because it is more similar to \cizoa{} than \cizob{}, as shown by the corresponding distances in Fig. \ref{fig:selected_mds}, the trajectories are directed to the former funnel. Another possible cause for this discrepancy is the small number of trajectories in our simulations, which was not able to resolve this rare event ($\SI{3.44}{\percent}$ according to Ref. \onlinecite{liu_new_2016}).

\begin{figure}[htbp]
    \centering
    \includegraphics[width=\linewidth]{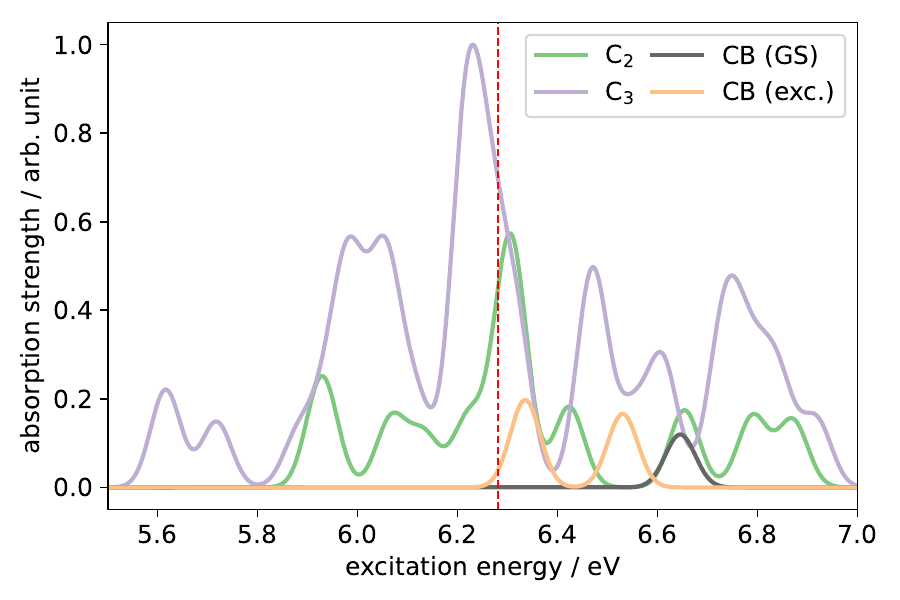}
    \caption{Broadened vertical absorption stick spectra with a Gaussian profile of width $\sigma = \SI{0.03}{\eV}$ of the transition $\mathrm{S_0}\to \mathrm{S_2}$ of the initial structures grouped by different photoproducts. The dashed red vertical line shows the median excitation energy.}
    \label{fig:ensemble_density}
\end{figure}

An additional consequence of the small number of trajectories we were able to run is that the data are not sufficient for the analysis of the correlation between the products and excited vibrational modes. Instead, we plotted the absorption strength by broadening the vertical absorption stick spectra of the transition $\mathrm{S_0}\to \mathrm{S_2}$ of the initial structures grouped by different photoproducts in Fig. \ref{fig:ensemble_density}. The \ccc{}:\cc{} ratio for the half of the trajectories with lower excitation energies is 4.2, higher than that for the half with higher excitation energies of 1.9. Therefore, we consider 4.2 to be an upper bound for excitations at \SI{200}{\nm}, which lies clearly in the lower energetic part after considering the calculated energies are redshifted, while the all-trajectory average of 2.8 acts like a lower bound.

\subsection{\label{subsec:results_electron_diffraction}Electron diffraction patterns}
\begin{figure*}[!ht]
    \centering
    \includegraphics[width=\linewidth]{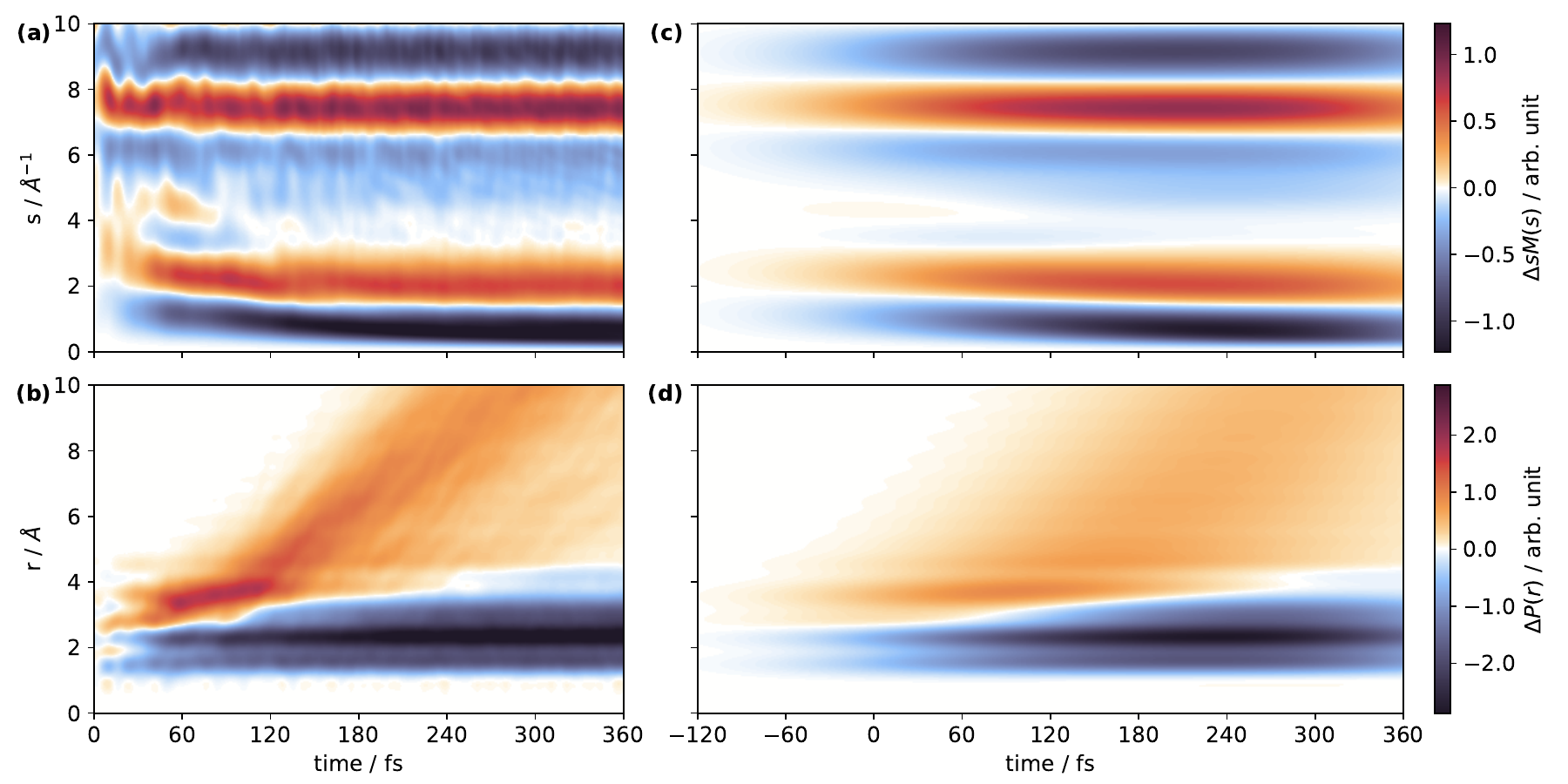}
    \caption{Simulated (a) time-dependent difference scattering intensity $\Delta s M(s;t)$ and (b) time-dependent difference pair distribution function $\Delta P(r;t)$ obtained from (a) by Eq. \eqref{eq:ged_difference_pdf} using $s_{\mathrm{min}}=\SI{0}{\per\angstrom}$, $s_{\mathrm{max}}=\SI{10}{\per\angstrom}$ and $\alpha = \SI{0.04}{\angstrom\squared}$. Panel (c) is obtained through folding of (a) with two Gaussians, which have the full-width at half-maximum of \SI{57}{\fs} and \SI{150}{\fs} to simulate the response from the pump and probe pulse present in an actual experimental setting. Panel (d) is obtained by applying the sine transform described in Eq. \eqref{eq:ged_difference_pdf} on (c).}
    \label{fig:gued}
\end{figure*}

To enable comparison between theoretical and experimental results, we have calculated the time-dependent difference scattering intensity $\Delta s M(s;t)$ and the time-dependent difference pair distribution function $\Delta P(r;t)$ according to Eq. \eqref{eq:ged_difference_signal} and \eqref{eq:ged_difference_pdf} for all trajectories using $s_{\mathrm{min}}=\SI{0}{\per\angstrom}$, $s_{\mathrm{max}}=\SI{10}{\per\angstrom}$ and $\alpha = \SI{0.04}{\angstrom\squared}$ as shown in Fig. \ref{fig:gued}a and \ref{fig:gued}b. The rotationally averaged signals defined in Eq. \eqref{eq:ged_intensity} were calculated using a script provided by Wolf \emph{et al.}\cite{wolf_diffraction_simulation_2021} For the reference signal $I_{\mathrm{mol}}(s;0)$, the optimized \cb{} structure was used. Every fifth frame in each trajectory was evaluated, giving a temporal resolution of \SI{1}{\fs}. We have provided geometries for all the trajectories in the spin-orbit coupled dynamics simulations in the supplementary material, so that the diffraction patterns can be easily regenerated using any other parameter set.

The negative signals centered at \SI{1.5}{\angstrom} and  \SI{2.3}{\angstrom} corresponds to the distance between two adjacent carbon atoms and that between one carbon atom and its second nearest neighbor in the cyclobutane ring. The former signal extends downwards to \SI{1.2}{\angstrom} through the distance between oxygen and carbonyl carbon (C\textsubscript{O}), whereas the latter signal extends upwards to \SI{3.5}{\angstrom} through the distances between oxygen and \textalpha- as well as \textbeta-carbon. The positive signals could be interpreted with the help of nonadiabatic dynamics simulations. The initial signals at \SI{2.8}{\angstrom} and \SI{3.6}{\angstrom} correspond to the \ch{C\textsubscript{O}\bond{semisingle}C\textsubscript{\textbeta{}}} and \ch{O\bond{semisingle}C\textsubscript{\textbeta{}}} distances, respectively, while the trajectories approach \cizoa{}. As the trajectory reaches \cizoa{}, these two distances become roughly the same and their signals merge to a broad trace centered at \SI{3.4}{\angstrom}. Since the distance between O and C\textsubscript{\textalpha{}} at the cleaved bond also has a similar value, this trace appears rather strong. Afterwards, the trajectories start to dissociate, creating the large area of positive signals, with the strongest part starting at around \SI{110}{\fs}. We also observe a minor trace of periodically occurring positive signal centered at \SI{0.9}{\angstrom}, which is caused by vibrations of the carbonyl group in \cb{} at shorter time delays and that of carbon monoxide or ethenone at later time delays. Its low intensity is explained by the out-of-phase oscillations across all trajectories. We also provide the time-dependent difference scattering intensity folded with two Gaussians, which have the full-width at half-maximum of \SI{57}{\fs} and \SI{150}{\fs}, to simulate the response from the pump and probe pulse present in an actual experimental setting. The resulted trace along with the obtained time-dependent difference pair distribution function are shown in Fig \ref{fig:gued}c and \ref{fig:gued}d.
A program visualizing the traces in Fig. \ref{fig:gued} and its contributions from each trajectory is provided in the supplementary material.

\begin{figure}[!ht]
    \centering
    \includegraphics[width=\linewidth]{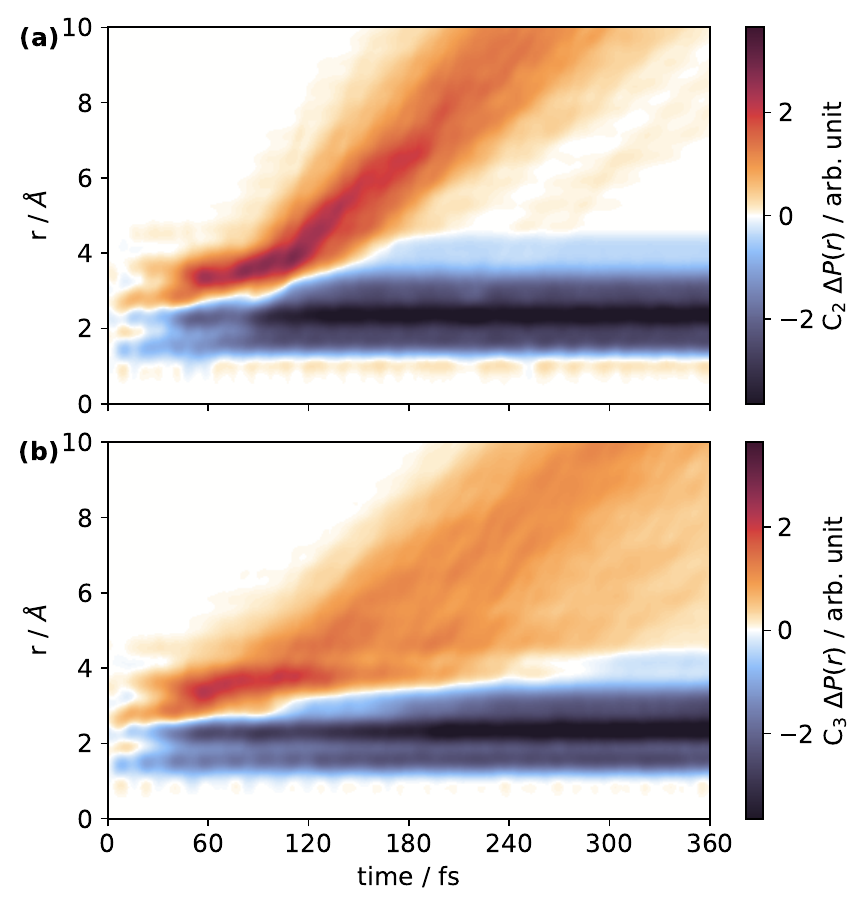}
    \caption{Time-dependent pair distribution function $\Delta P(r;t)$ averaged separately over trajectories following the (a) \cc{} and (b) \ccc{} pathway.}
    \label{fig:gued_c2c3}
\end{figure}

In order to understand the difference between trajectories following the \cc{} and \ccc{} pathway, we have plotted the time-dependent pair distribution functions averaged over each of these two classes of trajectories separately in Fig. \ref{fig:gued_c2c3}. While the negative signals do not differ much between these two classes, their positive signals are very different. The photodissociation leading to \cc{} products starts at around \SI{110}{\fs} for almost all trajectories, resulting in a sharp, line-shaped trace, while that leading to \ccc{} products is more spread out in time, thus forming a less intense, triangular-shaped trace. 

\section{\label{sec:conclusions}Conclusions and Outlook}

In this work, we have studied the ultrafast photodissociation of \cb{} starting in its 3s Rydberg state by employing trajectory surface hopping dynamics combined with the CASSCF(6,6)/MRCI method for the description of the electronic structure. The dynamics simulations were performed within a manifold of the lowest 3 singlet and 2 triplet states coupled by the spin-orbit interaction that has been treated with quasi-degenerate perturbation theory. Our simulations yielded two major dissociation pathways, with the photoproducts being either \ch{C2H4 + CH2CO} (\cc{} pathway) or \ch{"c-" C3H6 + CO} (\ccc{} pathway), consistent with the experimental findings.\cite{benson_photochemical_1942,blacet_photochemical_1957} We were also able to observe two secondary products, \ch{CH2 + CO} from the \cc{} pathway and \ch{C3H6} from the \ccc{} pathway, both of them previously reported in the literature.\cite{trentelman_193-nm_1990} The \ccc{}:\cc{} ratio was calculated to be 2.8 according to our simulations. With the help of multidimensional scaling, we could analyze the dissociation mechanism in terms of visited minimum-energy conical intersections.

Our simulations have shown that most trajectories reached their ground states after \SI{200}{\fs}, with dissociation events finished after \SI{300}{\fs}. Within the simulation time of \SI{360}{\fs}, only two trajectories remained in the S\textsubscript{1} state and one trajectory stayed as hot \cb{} in S\textsubscript{0}. Due to the low population of triplet states and the very similar branching ratio from dynamics simulations with only singlets included, we conclude that triplet states do not play any significant role in the photodissociation of \cb{} at this time scale. In order to establish direct link to future gas phase ultrafast electron diffraction experiments, we also provide a simulation of the time-resolved diffraction patterns. 




\section*{Data Availability Statement}
The data that support the findings of this study are available within the article and its supplementary material.


\bibliography{cyclobutanone}

\end{document}


\raggedbottom


\title{Supplementary Information: A CASSCF/MRCI Trajectory Surface Hopping Simulation of the Photochemical Dynamics and the Gas Phase Ultrafast Electron Diffraction Patterns of Cyclobutanone} 



\author{Xincheng Miao}
  \affiliation{Institut für Physikalische und Theoretische Chemie, Julius-Maximilians-Universität Würzburg, Emil-Fischer-Straße 42, 97074 Würzburg, Germany.}
\author{Kira Diemer}%
  \affiliation{Institut für Physikalische und Theoretische Chemie, Julius-Maximilians-Universität Würzburg, Emil-Fischer-Straße 42, 97074 Würzburg, Germany.}

\author{Roland Mitri\'{c}}
  \email{roland.mitric@uni-wuerzburg.de.}
  \affiliation{Institut für Physikalische und Theoretische Chemie, Julius-Maximilians-Universität Würzburg, Emil-Fischer-Straße 42, 97074 Würzburg, Germany.}


\date{\today}



{
\let\clearpage\relax
\maketitle
}
\tableofcontents
\vfill
\clearpage

\section{Active Space}

\begin{figure}[H]
    \centering
    \includegraphics[width=\linewidth]{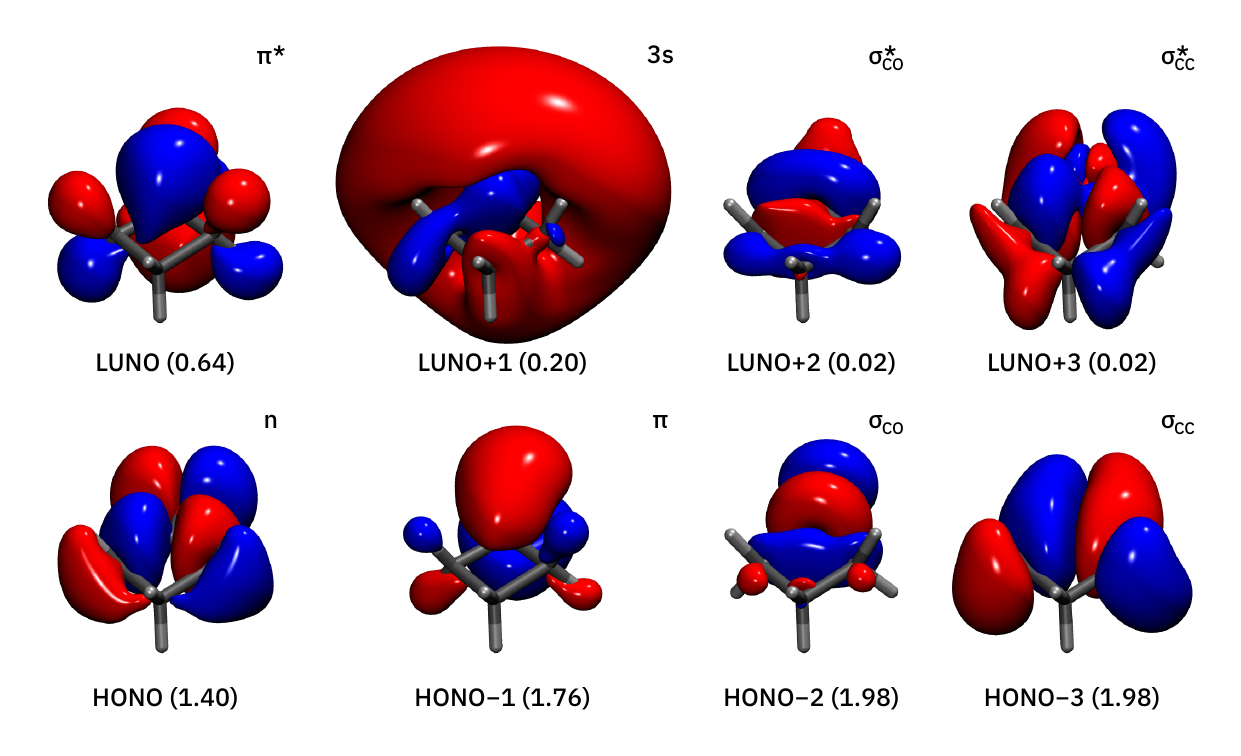}
    \caption{Natural orbitals from the SA(3/2)-CASSCF(8,8) calculation at the optimized ground state geometry of \cb{}. The occupation number of each orbital is given in the parenthesis and the character is shown in its upper right corner. An isovalue of 0.02 was chosen.}
    \label{fig:active_space_88}
\end{figure}

\begin{figure}[H]
    \centering
    \includegraphics[width=\linewidth]{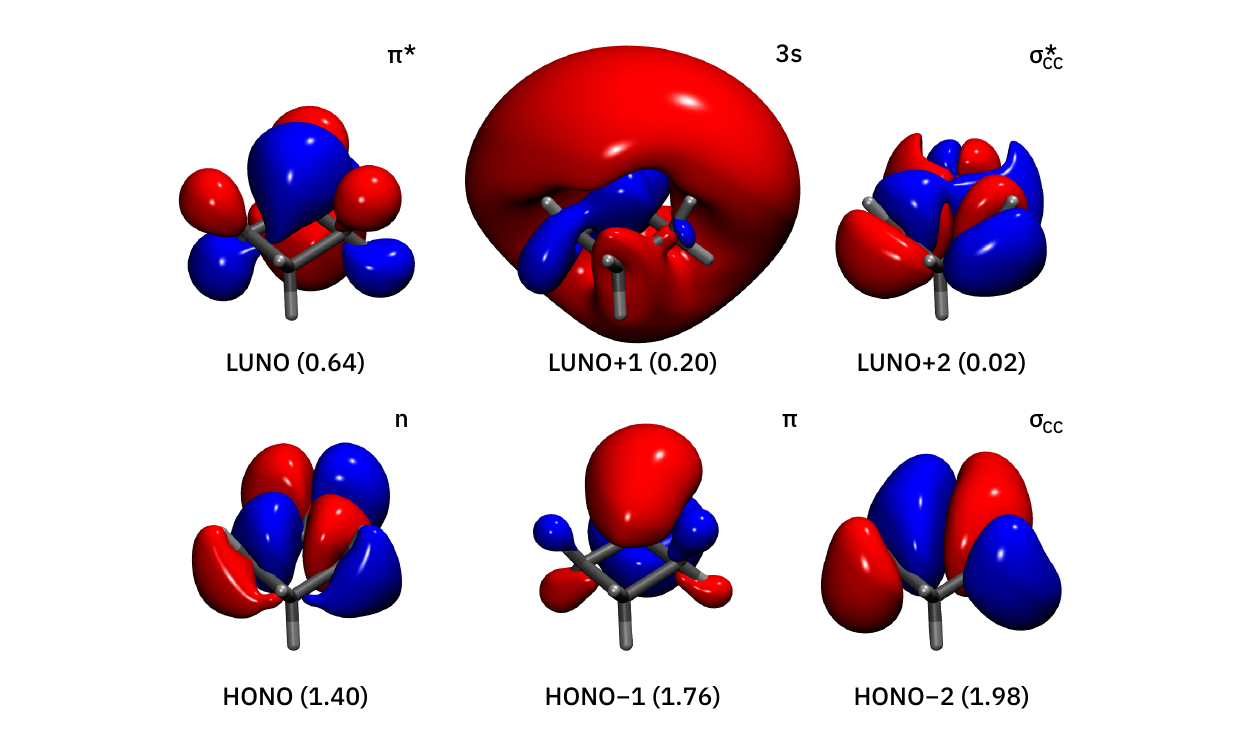}
    \caption{Natural orbitals from the SA(3/2)-CASSCF(6,6) calculation at the optimized ground state geometry of \cb{}. The occupation number of each orbital is given in the parenthesis and the character is shown in its upper right corner. An isovalue of 0.02 was chosen.}
    \label{fig:active_space_66}
\end{figure}

\section{Key Geometries in the Unit of \AA{}ngström}

\begin{table}[H]
    \begin{tabular}{lR{3cm}R{3cm}R{3cm}}
        \multicolumn{4}{l}{\cb{}} \\
        C &  -0.008788 &   0.071664 &   0.004068 \\
        C &   0.001577 &  -0.038233 &   1.532826 \\
        C &   1.554408 &   0.035720 &   1.475830 \\
        C &   1.512236 &   0.247860 &  -0.064695 \\
        H &  -0.481880 &   0.809035 &   2.019753 \\
        H &  -0.428931 &  -0.959870 &   1.923060 \\
        H &   1.992352 &   0.857874 &   2.037198 \\
        H &   2.043939 &  -0.889375 &   1.773070 \\
        H &   1.806735 &   1.242460 &  -0.400455 \\
        H &   2.026613 &  -0.494832 &  -0.673676 \\
        O &  -0.894687 &   0.015831 &  -0.843661 \\
\end{tabular}
\end{table}
\medbreak

\begin{table}[H]
    \begin{tabular}{lR{3cm}R{3cm}R{3cm}}
        \multicolumn{4}{l}{\cizoa{}} \\
        C &   1.010737 &   0.006022 &   0.071986 \\
        C &  -0.096641 &   1.014105 &   0.254932 \\
        C &  -1.301914 &   0.377685 &  -0.424521 \\
        C &  -1.345880 &  -0.992512 &   0.190205 \\
        H &  -0.296302 &   1.097074 &   1.326970 \\
        H &   0.206588 &   1.994100 &  -0.119825 \\
        H &  -2.199769 &   0.990856 &  -0.288317 \\
        H &  -1.121847 &   0.311415 &  -1.499890 \\
        H &  -1.477761 &  -1.062134 &   1.265668 \\
        H &  -1.650732 &  -1.852778 &  -0.394159 \\
        O &   1.829058 &  -0.839868 &   0.257026 \\
\end{tabular}
\end{table}
\medbreak

\begin{table}[H]
    \begin{tabular}{lR{3cm}R{3cm}R{3cm}}
        \multicolumn{4}{l}{\cizob{}} \\
        C &  -2.769349 &   0.239133 &  -0.171274 \\
        C &  -2.033686 &  -0.909735 &  -0.386568 \\
        C &  -4.886535 &  -1.284816 &  -0.051645 \\
        C &  -4.507988 &   0.076261 &   0.040815 \\
        H &  -0.963137 &  -0.834978 &  -0.520236 \\
        H &  -2.526037 &  -1.869701 &  -0.422600 \\
        H &  -5.072675 &  -1.749275 &  -1.012362 \\
        H &  -4.832259 &  -1.939871 &   0.809341 \\
        H &  -4.872166 &   0.711011 &  -0.758698 \\
        H &  -4.637550 &   0.524817 &   1.018972 \\
        O &  -2.455938 &   1.417007 &  -0.090316 \\
\end{tabular}
\end{table}
\medbreak

\begin{table}[H]
    \begin{tabular}{lR{3cm}R{3cm}R{3cm}}
        \multicolumn{4}{l}{\cizoc{}} \\
        C &  -1.032474 &   1.212233 &   1.772356 \\
        C &  -2.285296 &   1.742052 &   2.308551 \\
        C &  -4.314091 &  -0.501215 &  -0.000588 \\
        C &  -3.252725 &  -1.291326 &   0.001456 \\
        H &  -2.273097 &   2.603955 &   2.965840 \\
        H &  -3.231255 &   1.289352 &   2.052610 \\
        H &  -4.621928 &   0.042076 &   0.883152 \\
        H &  -4.922335 &  -0.363471 &  -0.884376 \\
        H &  -2.635825 &  -1.421917 &   0.880905 \\
        H &  -2.954248 &  -1.837691 &  -0.883638 \\
        O &   0.024109 &   1.753586 &   2.104237 \\
\end{tabular}
\end{table}
\medbreak

\begin{table}[H]
    \begin{tabular}{lR{3cm}R{3cm}R{3cm}}
        \multicolumn{4}{l}{\ciota{}} \\
        C &  -0.094351 &   0.093062 &  -0.076116 \\
        C &   0.081440 &  -0.039311 &   1.547540 \\
        C &   1.601180 &   0.019591 &   1.516913 \\
        C &   1.534817 &   0.240983 &   0.013363 \\
        H &  -0.451947 &   0.888042 &   1.785777 \\
        H &  -0.433024 &  -0.990706 &   1.705920 \\
        H &   1.989644 &   0.880753 &   2.042236 \\
        H &   2.087977 &  -0.898404 &   1.810362 \\
        H &   1.564790 &   1.277046 &  -0.342934 \\
        H &   1.824864 &  -0.555142 &  -0.677253 \\
        O &  -0.874043 &   0.034914 &  -0.825287 \\
\end{tabular}
\end{table}
\medbreak

\begin{table}[H]
    \begin{tabular}{lR{3cm}R{3cm}R{3cm}}
        \multicolumn{4}{l}{\ciotb{}} \\
        C &  -0.339551 &   0.028974 &  -0.236912 \\
        C &  -0.856282 &  -0.136431 &   1.112677 \\
        C &   1.586804 &   0.011961 &   0.991102 \\
        C &   1.122587 &   0.154734 &  -0.449193 \\
        H &  -1.033376 &   0.723759 &   1.752790 \\
        H &  -1.400362 &  -1.044086 &   1.369269 \\
        H &   1.785785 &   0.890738 &   1.589552 \\
        H &   1.958537 &  -0.936852 &   1.353395 \\
        H &   1.430945 &   1.107592 &  -0.890049 \\
        H &   1.526341 &  -0.643659 &  -1.080847 \\
        O &  -1.312242 &  -0.160993 &  -1.154253 \\
\end{tabular}
\end{table}

\section{Important Vibrational Modes in S\textsubscript{2}}

\begin{figure}[H]
    \centering
    \includegraphics[width=\linewidth]{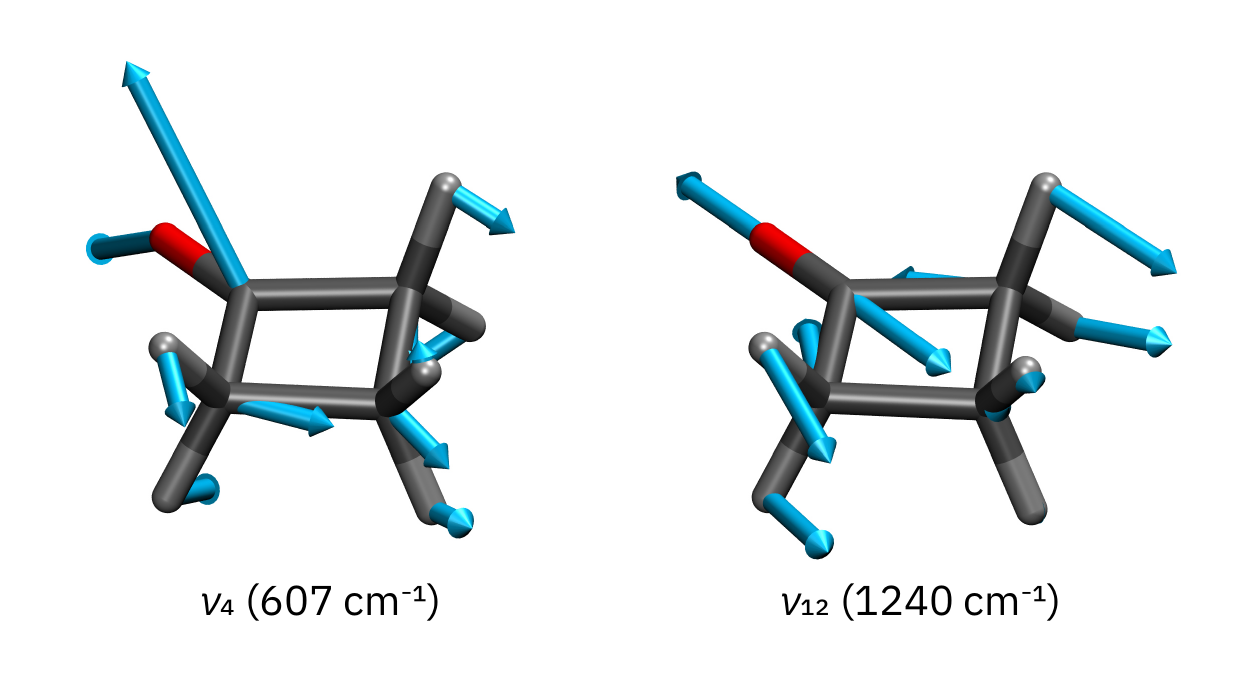}
    \caption{Important vibrational modes responsible for the vibronic structure in the absorption spectrum of \cb{} in S\textsubscript{2}. This is caused by the progression of the carbonyl stretching $\nu_{12}$ and its combination with the out-of-plane wagging of the carbonyl group $\nu_4$.}
    \label{fig:vibmodes}
\end{figure}

\section{State Energies from Nonadiabatic Dynamics}

\begin{figure}[H]
    \centering
    \includegraphics[width=\linewidth]{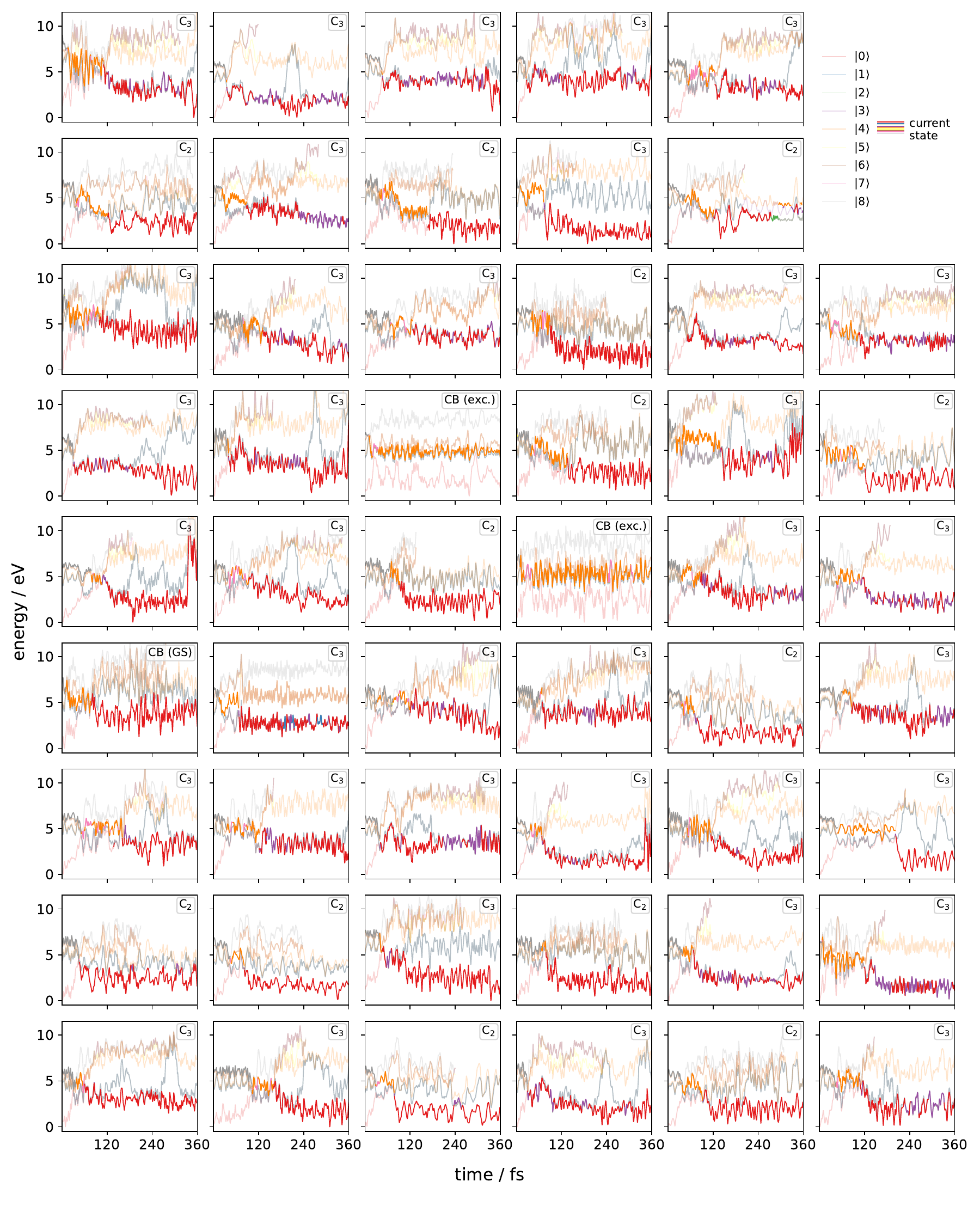}
    \caption{Time evolution of spin-adiabatic state energies for all trajectories. The current state of each trajectory is indicated by stronger coloring.}
    \label{fig:sad_energies}
\end{figure}
\clearpage

\begin{figure}[H]
    \centering
    \includegraphics[width=\linewidth]{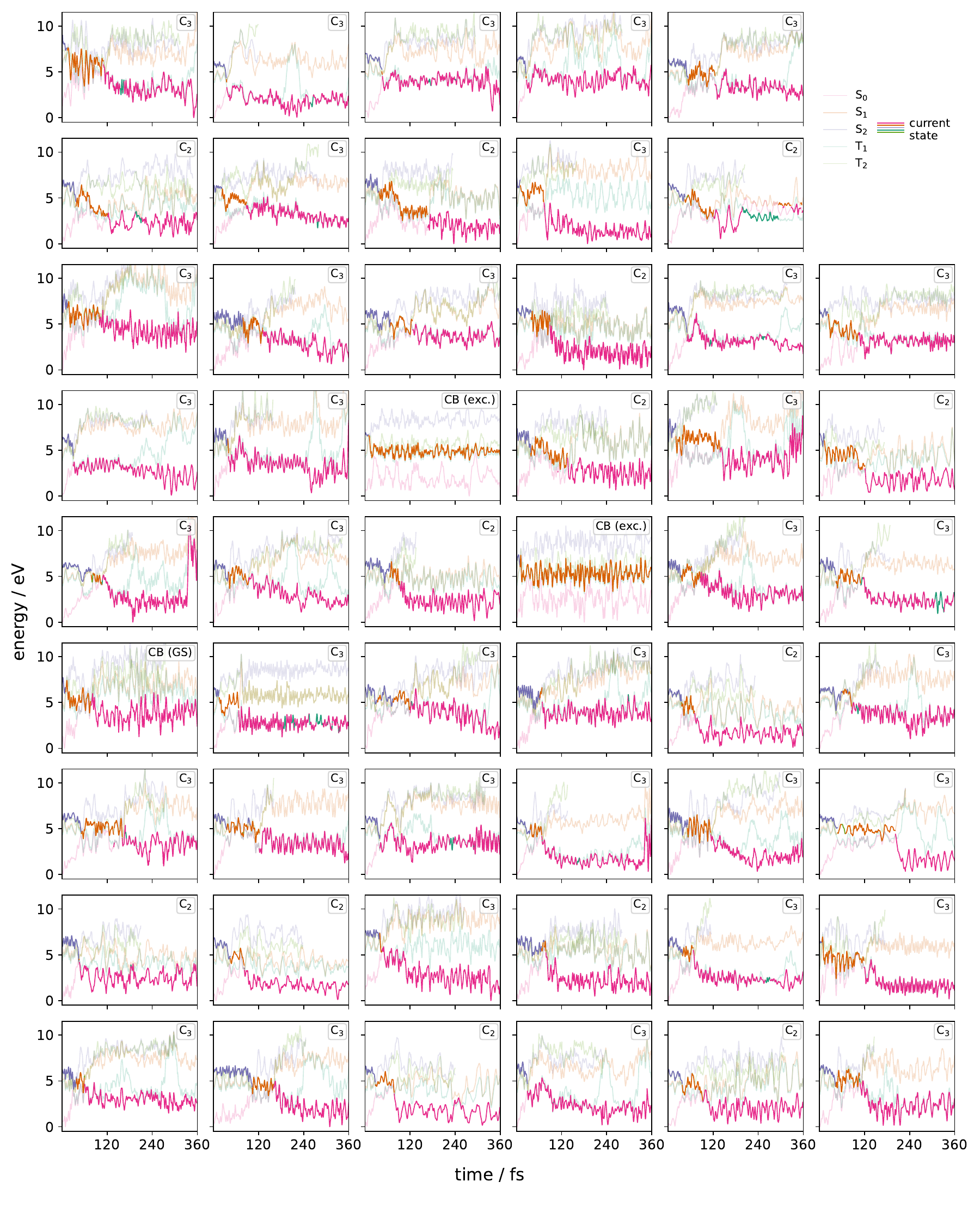}
    \caption{Time evolution of spin-diabatic state energies for all trajectories. The current state of each trajectory is indicated by stronger coloring.}
    \label{fig:sd_energies}
\end{figure}

\section{Rate Model}
The simulation results show that the molecule takes some time to reach the vicinity of conical intersections, where state transitions are enabled. Therefore, a delay time is required in addition to the lifetime to describe this delayed decay. By representing the molecular ensemble as a wavepacket, the probability density at the conical intersection will grow as the wavepacket travels towards it, corresponding to an increasing transition rate. Inspired by this consideration, we define the following rate function
\begin{equation}
    \label{eq:rate_function}
    k(t; t_\mathrm{g}, t_\mathrm{c}) = \begin{cases}
        0 & t < 0 \\
        \frac{t}{t_\mathrm{g} \cdot t_\mathrm{c}} & 0 \leq t \leq t_\mathrm{g} \\
        \frac{1}{t_\mathrm{c}} & t > t_\mathrm{g}
    \end{cases}\,,
\end{equation}
which pictures a linear growth of rate until $t_\mathrm{g}$, after which it assumes the maximum value of $k_{\mathrm{max}}=1/t_\mathrm{c}$. 

We then define the rate model as
\begin{subequations}
\begin{align}
    \label{eq:rate_model}
    \dot{P}_2(t) &= -k(t; t_{g,2}, t_{c,2}) P_2(t) \\
    \dot{P}_1(t) &= k(t; t_{g,2}, t_{c,2}) P_2(t) - k(t; t_{g,1}, t_{c,1}) P_1(t) \\
    \dot{P}_0(t) &= k(t; t_{g,1}, t_{c,1}) P_1(t)
\end{align}
\end{subequations}
with $P_i$ denoting the population in the $i$-th state. In the case of spin-diabatic populations, the $i$-th state represents $\mathrm{S}_i$, while for spin-adiabatic populations, $P_2$ and $P_1$ represent the population in the states $\ket{8}$ and $\ket{4}$, respectively. Since state $\ket{0}$ shows oscillations at later time due to its character change between singlet and triplet, which is not described by this simple rate model, its population was not included in the fit.

The mean rate is calculated as a weighted average of the rate function, where the weights are determined by the negative part of the population change, i.e.
\begin{subequations}
\begin{align}
    \label{eq:rate_weight}
    w_2(t) &= k(t; t_{g,2}, t_{c,2}) P_2(t) \\
    w_1(t) &= k(t; t_{g,1}, t_{c,1}) P_1(t)
\end{align}
\end{subequations}
and
\begin{equation}
    \label{eq:pop_mean_rate}
    \bar{k}_i = \frac{\int_0^\infty k(t; t_{g,i}, t_{c,i}) w_i(t) \du t}{\int_0^\infty w_i(t) \du t}\,.
\end{equation}

The mean lifetime of the $i$-th state is then defined as
\begin{equation}
    \label{eq:mean_lifetime}
    \bar{\tau}_i = \frac{1}{\bar{k}_i}\,.
\end{equation}

The parameters $t_{g,1}$, $t_{c,1}$, $t_{g,2}$ and $t_{c,2}$ were determined using least-square regression on the relative populations. The mean lifetimes $\bar{\tau}_1$ and $\bar{\tau}_2$ were then calculated after \eqref{eq:pop_mean_rate} and \eqref{eq:mean_lifetime} using the fitted parameters. The results are shown in table \ref{tab:pop_fit_res}.

\begin{table}[H]
    \setlength{\tabcolsep}{6pt}
    \centering
    \caption{The parameters of the rate model defined in eq. \eqref{eq:rate_model} fitted on populations in spin-adiabatic states $\ket{8}$ and $\ket{4}$, as well as in spin-diabatic states S\textsubscript{2}, S\textsubscript{1} and S\textsubscript{0}.}
    \label{tab:pop_fit_res}
    \begin{tabular}{rS[table-format=3.1]S[table-format=3.1]}
        \toprule[2.0pt]
        {parameter} & {spin-adiabatic} & {spin-diabatic}  \\ \midrule[2.0pt]
        $t_{g,1}$ / fs & 0.0 & 89.8 \\
        $t_{c,1}$ / fs & 55.2 & 47.9 \\
        $t_{g,2}$ / fs & 140.8 & 110.4 \\
        $t_{c,2}$ / fs & 7.2 & 9.3 \\ \midrule[0.4pt]
        $\tau_1$ / fs & 55.2 & 55.9 \\
        $\tau_2$ / fs & 25.4 & 25.6 \\ \bottomrule[2.0pt]
    \end{tabular}
\end{table}

\section{Photoproducts of Each Trajectory}
\begin{center}
\LTcapwidth=\textwidth
\setlength{\tabcolsep}{12pt}
\begin{longtable}{lll}
    \caption{Photoproducts of each trajectory after \SI{360}{\fs} from the dynamics simulations with spin-orbit coupling and that with only singlets.}
    \label{tab:photoproducts} \\
    
    \toprule[2.0pt] \multicolumn{1}{c}{traj.} & \multicolumn{1}{c}{spin-orbit simulation} & \multicolumn{1}{c}{singlet simulation} \\ \midrule[2.0pt]
    \endfirsthead
    
    \multicolumn{3}{c}%
    {\tablename\ \thetable{} -- continued from previous page} \\
    \toprule[2.0pt] \multicolumn{1}{c}{traj.} & \multicolumn{1}{c}{spin-orbit simulation} & \multicolumn{1}{c}{singlet simulation} \\ \midrule[2.0pt]
    \endhead
    
    \midrule[2.0pt] \multicolumn{3}{r}{{Continued on next page}} \\ \bottomrule[2.0pt]
    \endfoot
    
    \bottomrule[2.0pt]
    \endlastfoot

        00 & \ccc{} & \ch{C3H6 + CO} \\
        01 & \ccc{} & \ccc{} \\
        02 & \ccc{} & \ccc{} \\
        03 & \ccc{} & \ccc{} \\
        04 & \ccc{} & \ccc{} \\
        05 & \cc{} & \cc{} \\
        06 & \ccc{} & \ch{C2H4 + CH2 + CO} \\
        07 & \cc{} & \ccc{} \\
        08 & \ch{C3H6 + CO} & \ccc{} \\
        09 & \ch{C2H4 + CH2 + CO} & \cc{} \\
        
        10 & \ccc{} & \ccc{} \\
        11 & \ccc{} & \ccc{} \\
        12 & \ccc{} & \ccc{} \\
        13 & \cc{} & \cb{} (S\textsubscript{0}) \\
        14 & \ccc{} & \ccc{} \\
        15 & \ccc{} & \cb{} (S\textsubscript{1}) \\
        16 & \ccc{} & \ccc{} \\
        17 & \ccc{} & \ccc{} \\
        18 & \cb{} (S\textsubscript{1}) & \ccc{} \\
        19 & \cc{} & \ccc{} \\

        20 & \ccc{} & \ccc{} \\
        21 & \cc{} & \ccc{} \\
        22 & \ccc{} & \ccc{} \\
        23 & \ccc{} & \ccc{} \\
        24 & \cc{} & \cc{} \\
        25 & \cb{} (S\textsubscript{1}) & \cb{} (S\textsubscript{1})  \\
        26 & \ccc{} & \ccc{} \\
        27 & \ccc{} & \ccc{} \\
        28 & \cb{} (S\textsubscript{0}) & \ch{C3H6 + CO} \\
        29 & \ch{C3H5 + H + CO} & \ccc{} \\

        30 & \ccc{} & \cc{} \\
        31 & \ccc{} & \ccc{} \\
        32 & \cc{} & \ccc{} \\
        33 & \ccc{} & \ccc{} \\
        34 & \ccc{} & \ccc{} \\
        35 & \ccc{} & but-1-en-1-one \\
        36 & \ccc{} & \cc{} \\
        37 & \ccc{} & \cc{} \\
        38 & \ccc{} & \ccc{} \\
        39 & \ccc{} & \ccc{} \\

        40 & \cc{} & \ccc{} \\
        41 & \cc{} & \ccc{} \\
        42 & \ch{C3H6 + CO} & \ccc{} \\
        43 & \ccc{} & \ccc{} \\
        44 & \ccc{} & \ccc{} \\
        45 & \ccc{} & \ccc{} \\
        46 & \ccc{} & \cc{} \\
        47 & \ccc{} & \ccc{} \\
        48 & \cc{} & \cb{} (S\textsubscript{0}) \\
        49 & \ccc{} & \cc{} \\
        
        50 & \cc{} & \cc{} \\
        51 & \ccc{} & \ccc{} \\
        
\end{longtable}
\end{center}

\section{Molecular Descriptors and Dissimilarity Measures}
In order to guarantee translational and rotational invariance in our molecular representation, we utilized the strict upper triangular part of the molecular distance matrix, which is also inherently invariant under inversion. Given that our simulation deals with a dissociation reaction, we applied a clipping operation to the resulting vector with a maximum value of \SI{8}{Bohr}. In other words, the representation $\bm{p}$ is defined elementwise by
\begin{equation}
    \label{eq:mds_representation}
    p_{ab} = \min \left\{ \| \vec{R}_a - \vec{R}_b \|_2, \SI{8}{Bohr} \right\}
\end{equation}
for $a < b$, where $\vec{R}_a$ stands for the Cartesian coordinates of the $a$-th atom. This clipping ensures that the representation does not vary with the distances between fragments in the dissociated products. Furthermore, we define the weight $\bm{w}$ corresponding to the reduced mass of the two atoms involved in the particular distance, i.e.
\begin{equation}
    \label{eq:mds_weight}
    w_{ab} = \frac{M_a M_b}{M_a + M_b}\,,
\end{equation}
where $M_a$ stands for the molar mass of the $a$-th atom. Both the representation and the weight were then flattened to form the representation vector $p$ and the weight vector $w$.

Afterwards, we generated this representation for several nuclear permutation isomers (NPIs), which are structures with indices of identical nuclei permuted. To be specific, we have considered NPIs by interchanging the two \textalpha-carbons to ensure the C\textsubscript{s} symmetry of \cb{}, as well as those under permutation between \textalpha-carbon and \textbeta-carbon to account for the rotation of the cyclopropane ring against the carbon monoxide fragment observed in some trajectories. The set of representation vectors for all NPIs obtained at the geometry $i$ is denoted as $\{p^{i,n}\}_{n}$.

The dissimilarity $d_{ij}$ between two geometries $i$ and $j$ are calculated as the minimum weighted root-mean-square distance across all NPI-pairs from both geometries, i.e.
\begin{equation}
    \label{eq:mds_dissimilarity}
    d_{ij} = \min_{m,n} \frac{\| w \odot (p^{i,m} - p^{j,n}) \|_2}{\| w \|_2}\,,
\end{equation}
where $\odot$ denotes the elementwise product between 2 vectors. The multidimensional scaling analysis was performed using the dissimilary measure defined in eq. \eqref{eq:mds_dissimilarity} for each trajectory by selecting 181 geometries equidistantly, together with the 7 reference structures, namely \cb{}, \cc{}, \ccc{} and the five minimum-energy conical intersections.

\section{Additional Data and Visualization Scripts}
Additional data and visualization scripts are available on GitHub under \href{https://github.com/mitric-lab/Data_for_Cyclobutanone}{\url{https://github.com/mitric-lab/Data_for_Cyclobutanone}}.



%
%

%

